\documentclass[12pt,preprint]{aastex}

\newcommand{\eps}[1]{\mbox{log~$\epsilon$(#1)}} 
\newcommand\iso[2]{$^{\rm #1}$#2}

\def\eg{\mbox{e.g.}}
\def\etal{\mbox{\rm et al.}}

\def\rpro{\mbox{$r$-process}}
\def\spro{\mbox{$s$-process}}
\def\ncap{\mbox{$n$-capture}}
\def\teff{\mbox{T$_{\rm eff}$}}
\def\logg{\mbox{log~{\it g}}}
\def\vmicro{\mbox{$\xi_{\rm t}$}}
\def\kmsec{\mbox{km~s$^{\rm -1}$}}
\def\bd{\mbox{BD+17 3248}}
\def\cstwo{\mbox{CS 22892-052}}
\def\csthree{\mbox{CS 31082-001}}
\def\hdone{\mbox{HD 115444}}
\def\hdtwo{\mbox{HD 221170}}
\def\loggf{\mbox{$\log gf$}}

\shorttitle{Rare Earth Abundances}
\shortauthors{Sneden et al.}

\begin{document}

\title{New Rare Earth Element Abundance Distributions for the Sun and Five
       $r$-Process-Rich Very Metal-Poor Stars}

\author{
Christopher Sneden\altaffilmark{1,2},
James E. Lawler\altaffilmark{3},
John J. Cowan\altaffilmark{4}, \\
Inese I. Ivans\altaffilmark{5,6},
Elizabeth A. Den Hartog\altaffilmark{3}
}

\altaffiltext{1}{Department of Astronomy and McDonald Observatory, 
                 The University of Texas, Austin, TX 78712; 
                 chris@verdi.as.utexas.edu}

\altaffiltext{2}{INAF, Osservatorio Astronomico di Padova, Vicolo 
Osservatorio 5, I-35122 Padova, Italy}

\altaffiltext{3}{Department of Physics, University of Wisconsin,
                 Madison, WI 53706; 
                 jelawler@wisc.edu, eadenhar@wisc.edu}

\altaffiltext{4}{Homer L. Dodge Department of Physics and Astronomy, 
                 University of Oklahoma, Norman, OK 73019; 
                 cowan@nhn.ou.edu}

\altaffiltext{5}{The Observatories of the Carnegie Institution of 
                 Washington, Pasadena, CA 91101; iii@ociw.edu}

\altaffiltext{6}{Princeton University Observatory, Peyton Hall, 
                 Princeton, NJ 08544}

\begin{abstract}
We have derived new abundances of the rare-earth elements 
Pr, Dy, Tm, Yb, and Lu for the solar photosphere and for five
very metal-poor, neutron-capture \rpro-rich giant stars.
The photospheric values for all five elements are in good
agreement with meteoritic abundances.
For the low metallicity sample, these abundances have been combined
with new Ce abundances from a companion paper, and reconsideration of 
a few other elements in individual stars, to produce internally-consistent 
Ba, rare-earth, and Hf (56~$\leq$~Z~$\leq$~72) element distributions.
These have been used in a critical comparison between stellar
and solar \rpro\ abundance mixes.
\end{abstract}


\section{INTRODUCTION\label{intro}}

Early Galactic nucleosynthesis studies have been invigorated over the last 
decade by the discovery of many low metallicity halo stars with abundance 
distributions that depart significantly from that of our Solar System.
The neutron-capture elements (Z~$>$~30, hereafter \ncap) as a group 
exhibit particularly large star-to-star abundance variations 
with respect to Fe-peak elements.
For example, data from a number of surveys collected in Sneden, Cowan,
\& Gallino (2008)\nocite{sne08} show an abundance range in the rare-earth 
element Eu of at least $-$0.5~$\lesssim$ [Eu/Fe]~$\lesssim$ +2.0 
at metallicities [Fe/H]~$\lesssim$~$-$2.5\footnote{
We adopt the standard spectroscopic notation (Helfer, Wallerstein,
\& Greenstein 1959\nocite{hel59}) that for elements A and B,
[A/B] $\equiv$ log$_{\rm 10}$(N$_{\rm A}$/N$_{\rm B}$)$_{\star}$ $-$
log$_{\rm 10}$(N$_{\rm A}$/N$_{\rm B}$)$_{\odot}$.
We use the definition 
\eps{A} $\equiv$ log$_{\rm 10}$(N$_{\rm A}$/N$_{\rm H}$) + 12.0, and
equate metallicity with the stellar [Fe/H] value.};
see their Figure~14.

The \ncap\ abundances in the solar system and in most metal-rich Galactic 
disk stars arise from the combined effects of prior rapid and slow 
\ncap\ synthesis events (the ``\rpro'' and ``\spro'', respectively). 
The \ncap\ abundance patterns in low metallicity stars, however, vary widely. 
Examples have been found with element distributions that are consistent 
with the \rpro, the \spro, and a variety of mixes 
in between these two extremes.
These stars are thus natural test cases for \ncap\ nucleosynthesis
predictions.

Rigorous tests of \rpro\ and \spro\ theories require 
very accurate \ncap\ abundances in metal-poor stars.
Good abundance determinations result from effort on all fronts: 
acquisition of very high resolution, low noise spectra of the stars; 
construction of realistic model stellar atmospheres; analysis of the 
spectra with few limiting simplifications; and improvement in basic 
atomic and molecular data.
We have taken up the last consideration in the present series
of papers: 
Lawler, Bonvallet, \& Sneden (2001a\nocite{law01a}),
Lawler \etal\ (2001b\nocite{law01b}),
Lawler \etal\ (2001c\nocite{law01c}),
Den Hartog \etal\ (2003\nocite{den03}),
Lawler, Sneden, \& Cowan (2004\nocite{law04}),
Den Hartog \etal\ (2005\nocite{den05}),
Lawler \etal\ (2006\nocite{law06}),
Den Hartog \etal\ (2006\nocite{den06}),
Lawler \etal\ (2007\nocite{law07}),
Sobeck, Lawler, \& Sneden (2007\nocite{sob07}),
Lawler \etal\ (2008b\nocite{law08b}), and
Lawler \etal\ (2009\nocite{law09}).
We have concentrated most of our efforts on:
(a) improving the basic laboratory data for (mostly) rare-earth ionized 
species that are detectable in metal-poor stars; 
(b) applying these data to derive new solar spectroscopic abundances 
and comparing these photospheric values to solar-system meteoritic
data (Lodders 2003)\nocite{lod03};
and (c) extending the abundance analyses to a few well-studied low 
metallicity giants that are enriched in the products of the \rpro.
Our most recent study (Lawler \etal\ 2009\nocite{law09}) 
reports improved transition probabilities for 921 lines of \ion{Ce}{2}.
The present paper culminates this series with new solar and stellar
analyses of Pr, Dy, Tm, Yb, and Lu.
These elements all have good laboratory studies of their first
ions in the literature, but have not been systematically subjected 
to solar/stellar analyses in the same manner as have other rare earths.

In this paper we expand the standard definition of the rare-earth 
elements from the lanthanides (57~$\leq$ Z~$\leq$ 71) to include
two adjacent elements Ba (Z~=~56) and Hf (Z~=~72), and adopt 
the collective shorthand notation ``RE'' for them.
This broad definition covers a contiguous set of elements that have 
similar properties for stellar spectroscopy.
In particular, these elements have relatively low first ionization 
potentials, 5.2~eV~$\leq$ $IP$~$\leq$ 6.8~eV, and thus are almost
completely ionized in the solar photosphere and in the atmospheres 
of low-metallicity giant stars.
Their only detectable spectral features arise from their first ionized
species. 
Element groups in the Periodic Table immediately preceding the
REs (\eg, I, Xe, Cs) and following them (\eg, Ta, W, Re) 
have very different atomic properties. 
For various reasons traceable to very low abundances, Saha/Boltzmann 
energy level population effects, and/or lack of accessible transition
wavelengths, these elements just outside the RE group are 
inaccessible to most stellar spectroscopic detection efforts.

In \S\ref{obs} we review the solar and stellar spectroscopic data
and outline the abundance derivation methods.
Results for individual elements are given in \S\ref{newels}.
We summarize the total RE abundance sets for the solar
photosphere and \rpro-rich metal-poor giant stars in \S\ref{rareab}.
Finally, in \S\ref{discuss} we use the stellar RE abundance 
distributions in a critical examination of \rpro\ predictions.

\section{SPECTROSCOPIC OBSERVATIONS, REDUCTIONS AND ANALYSES\label{obs}}

For most of our stars, we analyzed the same high-resolution spectra
that have been used in previous papers of this series.
Additional descriptions of these stellar spectra can be found in
their original studies: 
\bd, Cowan \etal\ (2002)\nocite{cow02};
\cstwo, Sneden \etal\ (2003)\nocite{sne03};
\hdone, Westin \etal\ (2000)\nocite{wes00};
\hdtwo, Ivans \etal\ (2006)\nocite{iva06}; see also
Cowan \etal\ (2005)\nocite{cow05}.
The spectroscopic data sets employed in our analysis are 
summarized in Table~\ref{tab1}.
For each of the instrumental setups listed, we report the useful
wavelength range, and wavelength-dependent values of the signal-to-noise
$S/N$, resolving power $R$, and quality factor per resolution element
$F$ (sometimes also referred to as figure-of-merit),
at selected wavelengths $\lambda_{app}$.

Data reduction for the Keck and McDonald data have been detailed in
previous papers of this series and have largely relied on
IRAF\footnote{
IRAF is distributed by NOAO, which is operated by AURA,
under cooperative agreement with the NSF.}
and FIGARO.\footnote{
FIGARO is a part of the ``Starlink Project'', which is is now maintained
and being further developed by the Joint Astronomy Centre, Hawaii.}
For the recently acquired Magellan/MIKE data, we employed the MIKE
Pipeline software\footnote{
The MIKE Pipeline is available from the Carnegie
Observatories Software Repository at http://www.ociw.edu/Code/mike/}
(Kelson \etal\ 2000\nocite{kel00}, Kelson 2003\nocite{kel03}).
All of the data received final processing including continuum normalization
and telluric feature removal using SPECTRE \citep{fit87}.
Finally, for the solar analyses we employed the very high resolution 
($R$~= $\ge$~300,000), very high signal-to-noise (S/N~$\ge$ 1000) 
center-of-disk photospheric spectrum of Delbouille, Roland, 
and Neven (1973)\nocite{del73}\footnote{
Available at http://bass2000.obspm.fr/solar\_spect.php .}.

The abundance analyses used the same methods that have been described
at length in previous papers of this series.
Here we summarize the main points; the reader should consult Lawler 
\etal\ (2009\nocite{law09}) and references therein for details.

For each species, we begin with computations of relative strengths of 
all lines, in order to trim the often extensive laboratory line lists
to a set that might produce detectable absorption in the solar
photosphere and in our program stars.
Line absorption coefficients are proportional to products of oscillator 
strengths and absorber number densities.
In a standard LTE analysis, Boltzmann/Saha statistics describe the
populations of atoms in various ionization stages and electronic levels.
As discussed in \S\ref{intro}, the REs have low ionization
potentials, and thus exist almost completely as singly ionized species.
Saha corrections for other ionization states can be neglected.
Therefore the {\it relative} strength factors for ionized-species
RE elements can be approximated by 
$\log(\epsilon gf) - \theta \chi$, where $\epsilon$ is the elemental
abundance, $gf$ is the transition probability, 
$\theta$~=~5040/T is the reciprocal temperature, and $\chi$ is the
excitation energy.

\begin{figure}
\epsscale{0.9}
\plotone{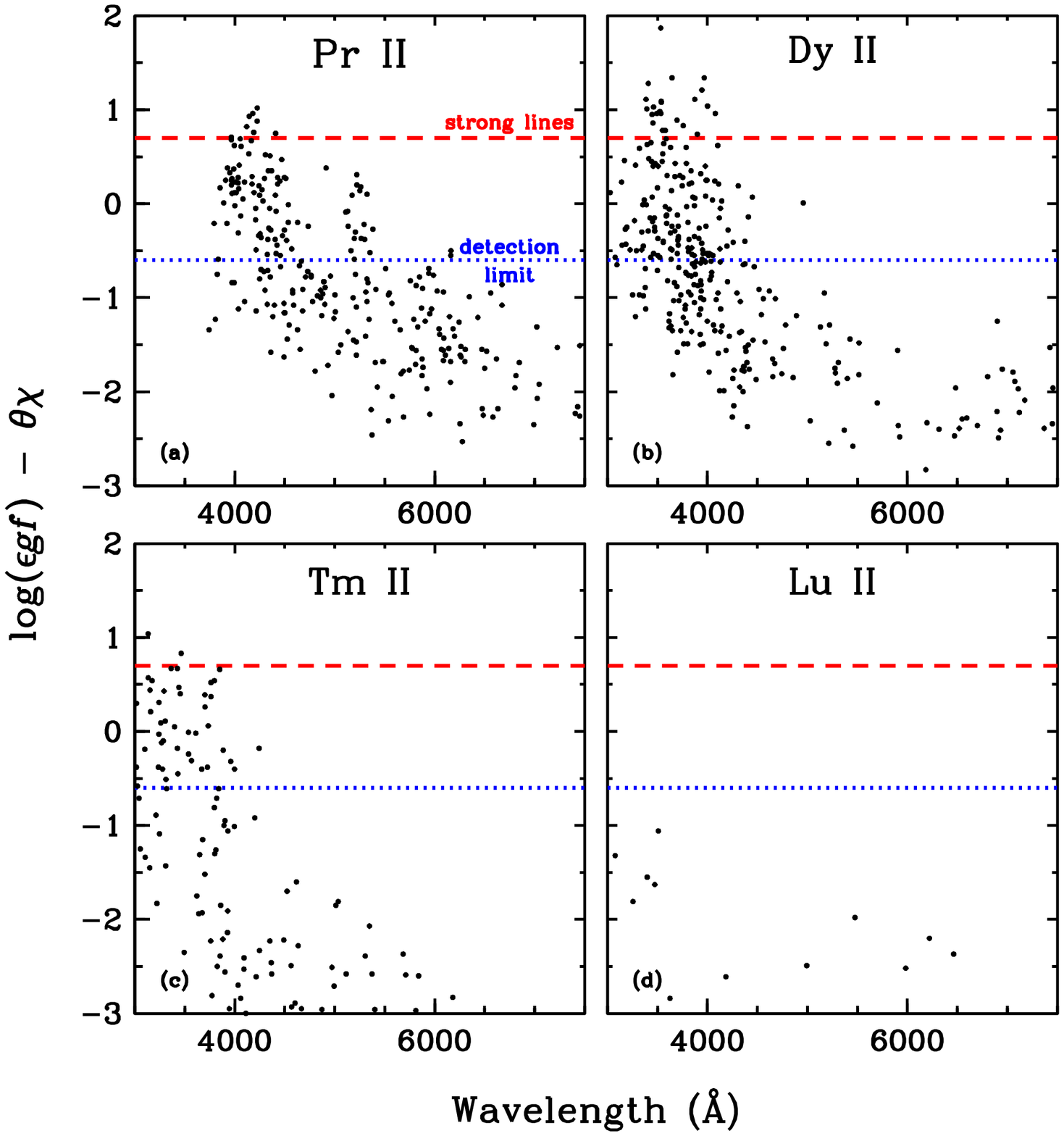}
\caption{
\label{f1} \footnotesize
Relative strength plots for \ion{Pr}{2}, \ion{Dy}{2}, \ion{Tm}{2}, and
\ion{Lu}{2} lines.
Dashed lines in each box denote the approximate lower strength limit
for strong lines, and dotted lines denote the lower limit for detectable
lines, as defined in the text.
For these plots the wavelength range has been restricted to 
$\lambda$~$>$~3000~\AA\ (the cutoff for ground-based spectra) and 
$\lambda$~$>$~7500~\AA\ (for lack of detectable lines of these species).
}
\end{figure}

Almost all easily detectable RE lines are of low excitation,
$\chi$~$\lesssim$~1~eV, so the relative line strengths are not
very sensitive to temperature.
Choosing $\theta$~=~1.0 as a rough mean of the solar and stellar 
reciprocal temperatures, and adopting approximate solar abundance values 
for each element under consideration, we computed relative strength
factors for \ion{Pr}{2}, \ion{Dy}{2}, \ion{Tm}{2}, and \ion{Lu}{2} lines,
using the laboratory data that will be discussed in the appropriate
subsections of \S\ref{newels}.  
We did not perform such computations for \ion{Yb}{2}, as it 
has only two very strong lines of interest for abundance
analyses (see \S\ref{ybtext}).
The results of this exercise are displayed in Figure~\ref{f1}.
With horizontal lines we mark the approximate minimum relative strength 
value for lines that can be considered ``strong''.
Such lines are those with evident saturation in their equivalent
widths (EWs), which for the Sun empirically is 
log($EW/\lambda$)~$\sim$~$-$5.3.  
We similarly mark the approximate strength value at which photospheric 
lines have log($EW/\lambda$)~$\sim$~$-$6.5, too weak to be routinely 
used in abundance solar abundance analyses.

Figure~\ref{f1} can be compared to similar plots for other 
RE elements in some of the previous papers of this series.
Some general remarks apply to all RE ions.
Most REs have complex energy structures, leading to large 
numbers of transitions.
Their relative strength factors increase with decreasing wavelength;
these usually are transitions from the lowest energy levels with the
largest \loggf\ values.
The most fertile regime for RE transitions is the near-UV
domain, $\lambda$~$<$~4000~\AA.
Unfortunately, the strong-line density of all species increases 
in this wavelength range, and many promising RE transitions
are hopelessly blended with (usually) Fe-peak lines.
Finally, as is evident in Figure~\ref{f1}, very few RE ions
have detectable transitions in the yellow-red ($\lambda$~$>$~5000~\AA)
spectral region of the solar spectrum.
Comments on the line strengths of individual species will be given
in \S\ref{newels}.
These same strength factors turn out to work reasonably well for 
the \rpro-rich giant stars.  
Their combination of cooler temperatures, more extended atmospheres,
metal poverty, and enhanced \ncap\ abundances yields line strengths
that are similar to or somewhat larger than those for the Sun.

We eliminated lines with relative strength factors that fell below 
the probable detection limits, and searched solar and stellar spectra
for the remaining lines.
In this effort we employed the large Kurucz (1998)\nocite{kur98}\footnote{
Available at http://cfaku5.cfa.harvard.edu/}
line list, the solar line identifications of Moore, Minnaert, \& Houtgast
(1966)\nocite{moo66}, and the observed spectra described above.
With these resources we were able to discard many additional lines
that proved to be too weak and/or too blended to be of use either for the
Sun or for the \rpro-rich stars.

We then constructed synthetic spectrum lists for small spectral regions 
(4--6~\AA) surrounding each promising candidate line.
These lists were built beginning with the Kurucz (1998)\nocite{kur98}
atomic line database. 
We updated the \ncap\ species transition probabilities with results from 
this series of papers, including the laboratory data cited below for 
Pr, Dy, Tm, Yb, and Lu.
We also used recently published \loggf\ values for \ion{Cr}{1} 
(Sobeck, Lawler, \& Sneden 2007)\nocite{sob07} and \ion{Zr}{2}
(Malcheva \etal\ 2006).\nocite{mal06}
Lines missing from the Kurucz database but listed in the laboratory
studies or in the Moore \etal\ (1966)\nocite{moo66} solar line atlas
were added in.
In spectral regions where molecular absorption is important, we used
the Kurucz data for OH, NH, MgH, and CN, and Plez (private communication)
data for CH.

We iterated the transition probabilities through repeated trial spectrum 
syntheses of the solar photosphere (and sometimes one of the \rpro-rich
giant stars).  
For the Sun, as in previous papers of this series we adopted the 
Holweger \& M\"uller (1974)\nocite{hol74} empirical model photosphere,
and computed the synthetic spectra with the current version of 
Sneden's (1973)\nocite{sne73} LTE, 1-dimensional (1D) line analysis code MOOG.
In these trial syntheses, no alterations were made to the lines with 
good laboratory \loggf's.
On occasion, obvious absorptions without plausible lab or solar 
identifications were arbitrarily defined to be \ion{Fe}{1} lines with 
excitation energies $\chi$~=~3.5~eV and \loggf\ values 
to match the photospheric absorption.
We discarded all candidate RE lines that proved to be seriously
blended with unidentified contaminants.

Final solar abundances for each line were determined through matches 
between the Delbouille \etal\ (1973)\nocite{del73} photospheric 
center-of-disk spectra and the empirically-smoothed synthetic spectra.
The same procedures were applied to the observed stellar spectra 
(Table~\ref{tab1}) and synthetic spectra generated with the
model atmospheres whose parameters and their sources are given in 
Table~\ref{tab2}.

\section{ABUNDANCES OF PR, DY, TM, YB, AND LU\label{newels}}

In this section we discuss our abundance determinations of 
elements Pr, Dy, Tm, Yb, and Lu in the Sun and 
the \rpro-rich stars.
Tables~\ref{tab3}, \ref{tab4}, and \ref{tab5}
contain the mean abundances in the solar photosphere and in the 
\rpro-rich low metallicity giants for these elements and for other 
REs that have been analyzed in previous papers of this series.
The full suite of elements will be discussed in \S\ref{rareab}.
Table~\ref{tab3} also gives estimates of \rpro\ abundance components
in solar-system meteoritic material.
These data will be discussed in more detail in \S\ref{discuss}.

\subsection{Praseodymium\label{prtext}}

Pr ($Z$~=~59) has one naturally-occurring isotope,
\iso{141}{Pr}. 
The \ion{Pr}{2} spectrum has been well studied in the laboratory,
with transition probabilities reported by Ivarsson \etal\ 
(2001; hereafter Iv01)\nocite{iva01}, Bi\'emont \etal\ 
(2003)\nocite{bie03} and Li \etal\ (2007; hereafter Li07), as
well as numerous publications on its wide hyperfine structure.
We will consider the hfs data in more detail in the Appendix.

\begin{figure}
\epsscale{0.9}
\plotone{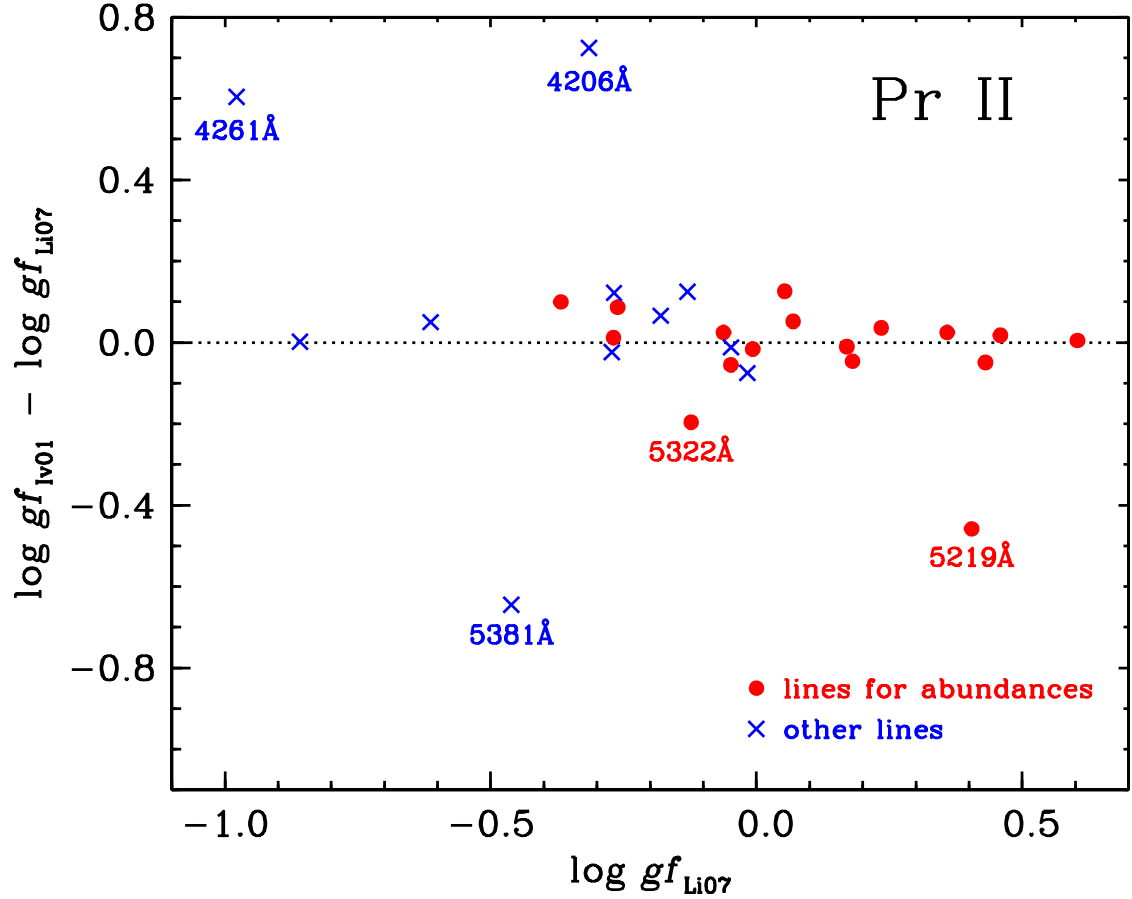}
\caption{Differences between Ivarsson \etal\ (2001; Iv01) and 
Li \etal\ (2007; Li07) \ion{Pr}{2} \loggf\ values
plotted as a function of wavelength.
As indicated in the figure legend, red dots denote transitions employed
in our solar/stellar analyses, and blue $\times$ symbols denote other 
lines in common between Iv01 and Li07.
\label{f2} \footnotesize
}
\end{figure}

We adopted Li07 as our primary transition probability source.
This is the most recent and largest set, 260 lines, 
of purely experimental measurements (Li07 combined their own branching
fractions with previously published lifetimes).
Iv01 also conducted a smaller \ion{Pr}{2} lab study, 
reporting \loggf values for 31 lines.
However, their list includes four lines not published by Li07.
Therefore we considered both Li07 and Iv01 data sets in our abundance
determinations.
In Figure~\ref{f2} we plot the differences between individual
Iv01 and Li07 \loggf\ values, using different symbols to distinguish
those lines employed in our abundance analyses from those that
proved to be unsuitably weak or blended.
There is generally good agreement: ignoring the five obviously
discrepant lines that are labeled by wavelength in the figure,
the mean difference is
$<\loggf_{Iv01} - \loggf_{Li07}>$ = +0.03~$\pm$~0.01
($\sigma$~=~0.06, 23 lines). 
Comments on individual lines in common are given below.
Bi\'emont \etal\ (2003)\nocite{bie03} also published \loggf's 
for 150 \ion{Pr}{2} lines. 
However, their values were determined by combining experimental 
\ion{Pr}{2} lifetimes and theoretical branching fractions, which are 
very difficult to compute for the complex RE atomic structures 
(\eg, Lawler \etal\ (2008a).\nocite{law08a}

Moore \etal\ (1966)\nocite{moo66} give 21 \ion{Pr}{2} identifications
for the solar spectrum.
However, most of them are very weak and/or blended.
An early study by Bi\'emont \etal\ (1979)\nocite{bie79} has a good 
discussion of the benefits and disadvantages of many of these lines for 
photospheric abundance work.
They used nine lines to derive \eps{Pr}$_\odot$~= 0.71~$\pm$~0.08,\footnote{
Throughout this paper we will use the subscript symbol $\odot$ to
indicate solar photospheric values, and the subscript $met$ to indicate 
solar system meteoritic values from Lodders (2003)\nocite{lod03}.}
with individual lines contributing to the average with different weights.
Only three of these lines were considered to be high-quality ones.
More recently, Ivarsson, Wahlgren, \& Ludwig (2003)\nocite{iva03}
employed synthetic/observed spectral matches to suggest 
\eps{Pr}$_\odot$~= 0.4~$\pm$~0.1, more than a factor of two smaller
than the meteoritic value of \eps{Pr}$_{met}$~= 0.78~$\pm$~0.03.

\begin{figure}
\epsscale{0.85}
\plotone{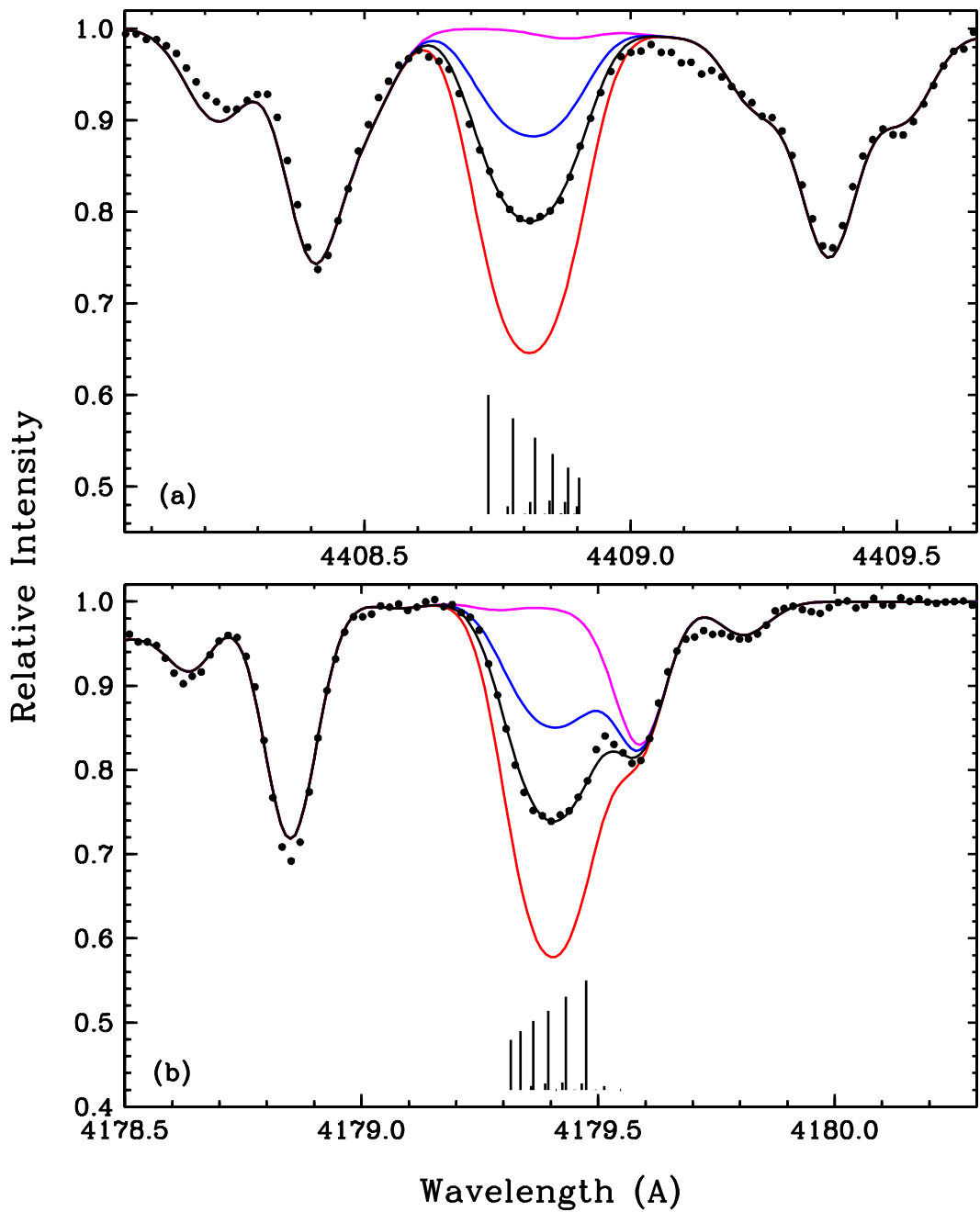}
\caption{Observed and synthetic spectra in CS~31082-001 of two strong 
\ion{Pr}{2} lines with wide hyperfine structure.
In each panel, the points represent the observed spectrum.
The magenta line is the spectrum computed with no contribution
from \ion{Pr}{2}; the black line is the best-fitting synthesis (with the
Pr abundance given in Table~\ref{tab6}; and the red and blue lines
are the syntheses computed with Pr abundances altered by $\pm$0.3~dex
from the best value.
Vertical lines have been drawn at the bottom of each panel to
indicate the wavelengths and {\it relative} strengths (arbitrary
overall normalization) of the hyperfine
components that comprise the \ion{Pr}{2} transitions.
\label{f3} \footnotesize
}
\end{figure}

We searched for useful \ion{Pr}{2} lines in the solar spectrum by first
identifying them in \csthree, which is the most extreme \rpro-rich 
metal-poor star of our sample: [Fe/H]~= --2.9, [Eu/Fe]~= +1.6
(Hill \etal\ 2002\nocite{hil02}).
This star's low metallicity and large [\ncap/Fe-peak] abundance ratios
combine to yield many strong (and often essentially unblended) 
candidate transitions.
Inspection of the \csthree\ spectrum yielded 43 lines from Li07 and an 
additional 3 lines from Iv01 that merited abundance consideration.
Preliminary synthetic spectrum calculations suggested that 13 of 
these candidate lines were either too weak or too blended
in both \csthree\ and the Sun.
The wide hyperfine structure of all prominent \ion{Pr}{2} lines made this 
exercise much easier than it would be in searches for lines with no hfs.
In Figure~\ref{f3} we illustrate this point with synthetic and
observed spectra of the strong 4408.8 and 4179.4~\AA\ transitions.
Visual inspection of the \ion{Pr}{2} profiles suggests that their 
full-width-half-maxima are FWHM~$\simeq$~0.4~\AA, while observed and
synthetic profiles of single lines (\eg, 4178.86~\AA\ \ion{Fe}{2} and 
4179.59~\AA\ \ion{Nd}{2}) have FWHM~$\simeq$~0.25~\AA\

Wavelengths of the remaining useful \ion{Pr}{2} lines are given in 
Table~\ref{tab6}, along with their excitation energies and the
Li07 and Iv01 transition probabilities.
In Figure~\ref{f2} one sees five lines with large \loggf\ 
discrepancies between these studies.
Three of the lines were not involved for our abundance studies and so
we cannot comment further on them.
Li07 caution that 5219.1~\AA\ is blended on their spectra.
We adopted the Iv01 value for this line.
Finally, the difference between Iv01 and Li07 for 5322.8~\AA\ is 0.2~dex,
but abundances derived with the Li07 \loggf\ proved to be consistent with
those from other Pr lines.

\begin{figure}
\epsscale{0.85}
\plotone{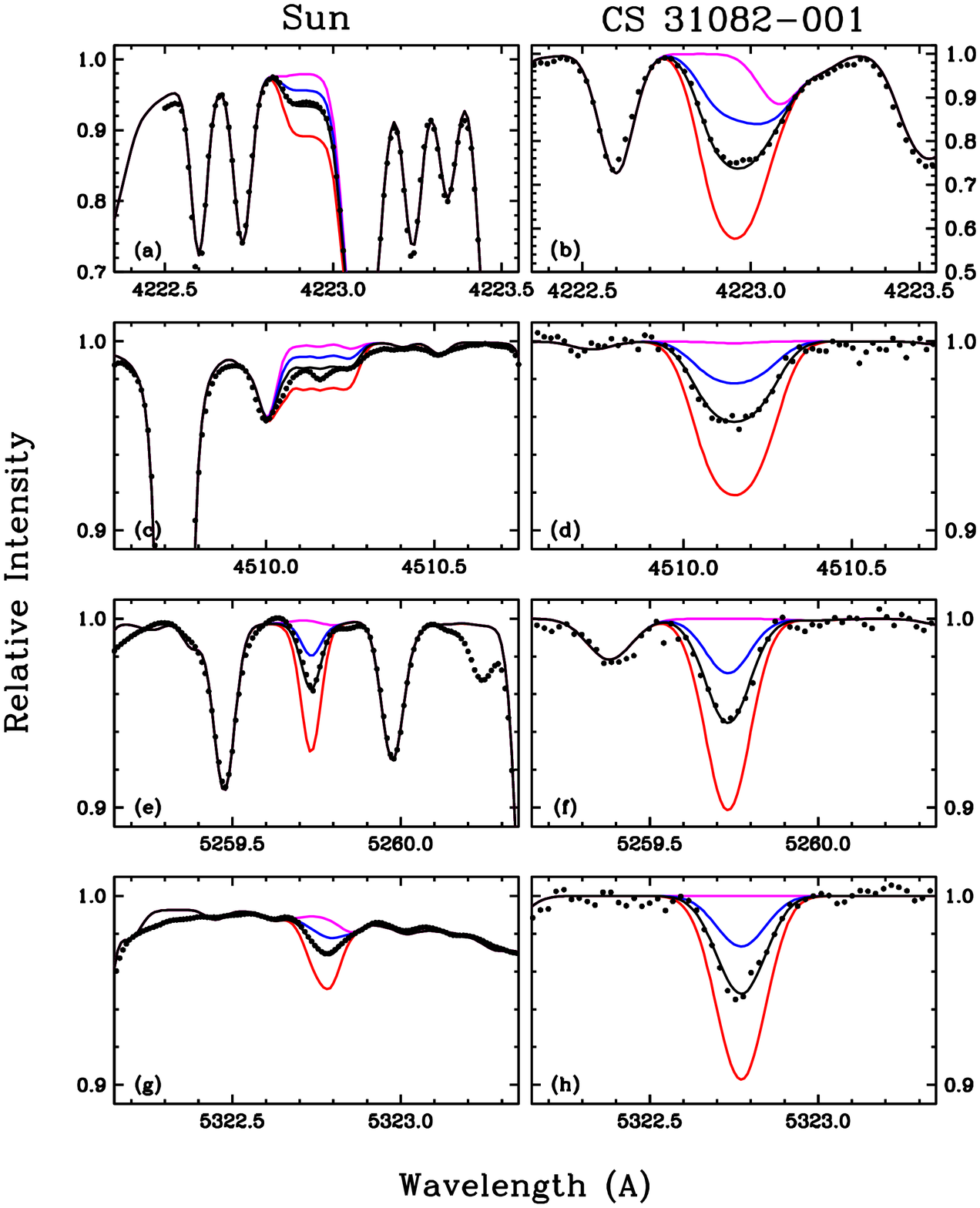}
\caption{Observed and synthetic spectra of the Sun (left-hand 
panels (a), (c), (e), (g)) and CS~31082-001 (right-hand panels
(b), (d), (f), (h)) for the four \ion{Pr}{2} lines that contribute 
to the solar abundance estimate.
In each panel, the points represent the observed spectrum.
The magenta line is the spectrum computed with no contribution 
from \ion{Pr}{2}; the black line is the best-fitting synthesis (with the
Pr abundance given in Table~\ref{tab6}); and the red and blue lines
are the syntheses computed with Pr abundances altered by $\pm$0.3~dex
from the best value.
The solar spectrum is that of Delbouille \etal\ (1973), but sampled at
a wavelength step size of 0.01~\AA\ for display purposes.
\label{f4} \footnotesize
}
\end{figure}

We calculated solar photospheric synthetic spectra for all the 
\ion{Pr}{2} lines of Table~\ref{tab6}.
We found, as have the previous studies cited above, that there are 
few useful solar Pr abundance indicators.
Our final value was based on five lines (Table~\ref{tab6}).
We show the synthetic/observed photospheric spectrum matches for
four of these lines in left-hand panels (a), (c), (e), and (g) of
Figure~\ref{f4}, contrasting their appearance in right-hand 
panels (b), (d), (f), and (h) for \csthree.
We do not include the 5219.1~\AA\ line in Figure~\ref{f4} because 
it was too weak in the spectrum of \csthree\ to analyze in that star.
Note that Li07 transition probabilities were used for the 4222.9,
4510.1, and 5322.8~\AA\ lines and Iv01 values for the 5219.1 
and 5259.7~\AA\ lines.
However, consistent abundances from all five lines were derived: the
mean value (Table~\ref{tab3}) is \eps{Pr}$_\odot$~= 0.76~$\pm$~0.02 
(sigma~=~0.04).
Our new photospheric abundance is in good agreement with
the meteoritic and the Bi\'emont \etal\ (1979)\nocite{bie79}
photospheric abundances that were quoted above.

\begin{figure}
\epsscale{0.85}
\plotone{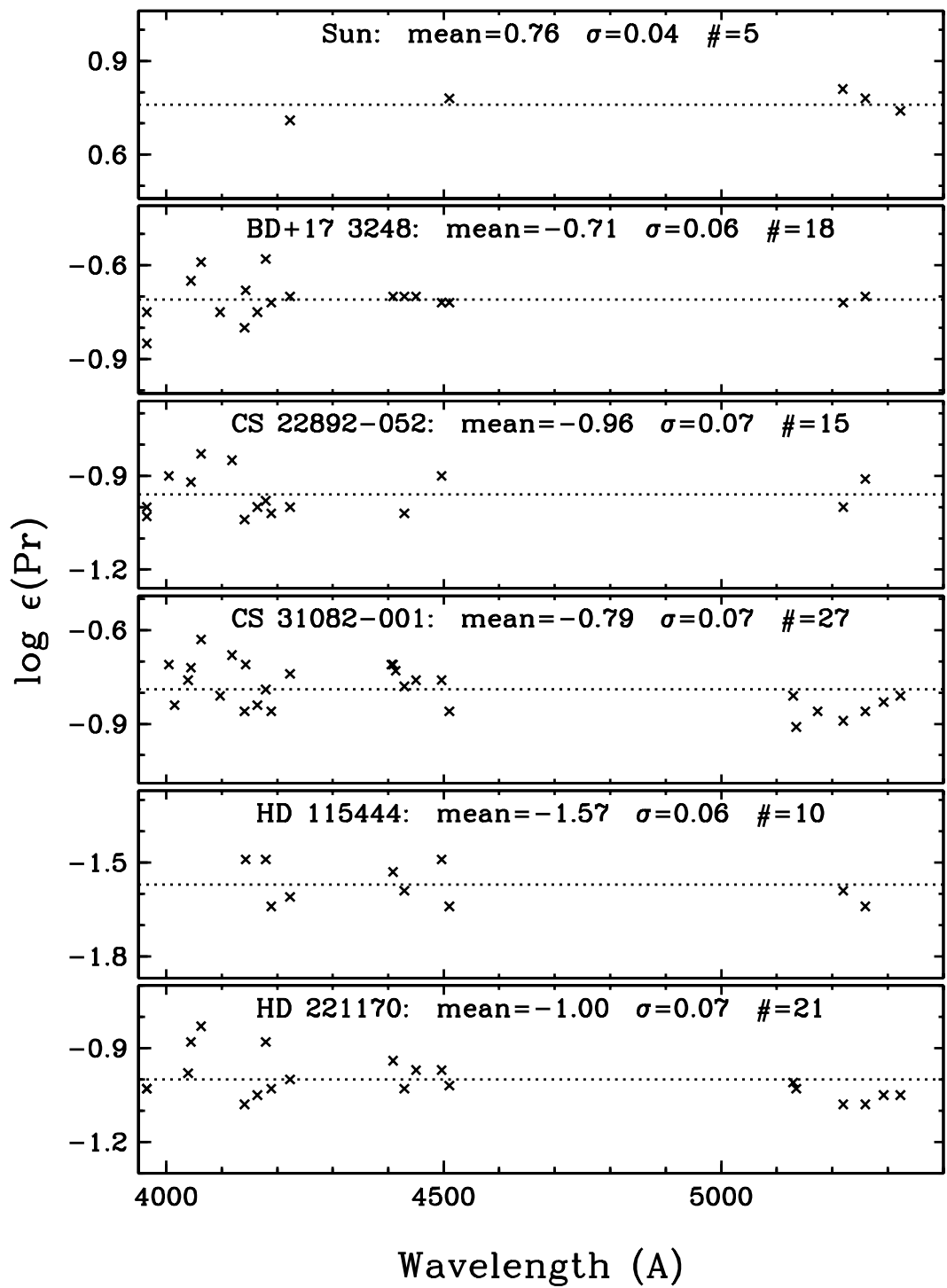}
\caption{Derived Pr abundances for the Sun and the \rpro-rich
low metallicity stars plotted as functions of wavelength.
The abundance range shown for each star is 0.6~dex, and is centered
vertically on the mean abundance, which is indicated with a dotted line.
The legend of each panel records this abundance mean, along with the
sample standard deviation and number of transitions used.
\label{f5} \footnotesize
}
\end{figure}

For the \rpro-rich low metallicity stars we derived abundances 
from 10--27 \ion{Pr}{2} lines (Table~\ref{tab6}).
We plot the individual line abundances for these stars and the Sun
as functions of wavelength in Figure~\ref{f5}, with their summary 
abundance statistics in the panel legends.
In each case the line-to-line scatter was small, $\sigma$~$\simeq$~0.06,
and we found no significant abundance trends with wavelength,
excitation energy (the range in this quantity is only
$\simeq$1~dex), or \loggf.

\subsection{Dysprosium\label{dytext}} 

Dy ($Z$~=~66) has seven naturally-occurring isotopes,
five of which contribute substantially to its solar-system abundance:
\iso{156,158}{Dy}, $\ll$1\%, 
\iso{160}{Dy}, 2.34\%;
\iso{161}{Dy}, 18.91\%; 
\iso{162}{Dy}, 25.51\%; 
\iso{163}{Dy}, 24.9\%; and 
\iso{164}{Dy}, 28.19\% (Lodders 2003\nocite{lod03}).
The atomic structure of \ion{Dy}{2} is complex, leading to a rich
spectrum of transitions arising from low-excitation energy levels.
This species has been well-studied in the laboratory recently, with
published transition probabilities by Kusz (1992)\nocite{kus92},
Bi\'emont \& Lowe (1993)\nocite{bie93}, and Wickliffe, Lawler,
and Nave (2000)\nocite{wic00}.
The Wickliffe \etal\ study contains a detailed comparison of their 
transition probabilities with those of Kusz and Bi\'emont \& Lowe
(as well as earlier investigations), and will not be repeated here.

We adopted the Wickliffe \etal\ (2000)\nocite{wic00} \loggf\ values,
as in our earlier analyses of the \rpro-rich stars.
Those studies (\eg, Ivans \etal\ 2006\nocite{iva06} for \hdtwo, 
and Sneden \etal\ 2003\nocite{sne03} for 
\cstwo) performed extensive searches for promising \ion{Dy}{2} lines.
However, the Dy abundances reported in those papers were derived
from both EW matches and synthetic spectrum calculations.
Therefore, to be internally consistent in our new analyses, we began
afresh with new solar \ion{Dy}{2} identifications and new synthesis
line lists for each chosen feature.
In principle \ion{Dy}{2} lines should have both isotopic wavelength
splitting and (for \iso{161,163}{Dy}) hyperfine substructure.
We inspected the profiles of many of the strongest lines 
appearing in National Solar Observatory (NSO) Fourier Transform
Spectrometer (FTS) laboratory \ion{Dy}{2} spectra.
Some line substructure is present in each line.  
However, the components that are shifted away from the line centers 
are always very weak ($\lesssim$10\% of central intensities),
and the full widths near profile baselines are $\sim$0.05~\AA.
For all lines, FWHM~$\sim$~0.02~\AA\ in the lab spectra.
These widths are substantially smaller than the measured solar and 
stellar spectrum line widths.
Therefore we treated all \ion{Dy}{2} lines as single features.

There are many candidate lines, as indicated by their relative
strength values shown in panel (b) of Figure~\ref{f1}.
Solar Dy abundances could be determined from 13 of these transitions.
The resulting mean photospheric abundance (Table~\ref{tab3}) is
\eps{Dy}$_\odot$~= +1.13~$\pm$~0.02 ($\sigma$~= 0.06).
This value is in excellent agreement with the meteoritic abundance, 
\eps{Dy}$_{met}$~= +1.13~$\pm$~0.04 and with the 
Kusz (1992)\nocite{kus92} photospheric abundance, 
\eps{Dy}$_\odot$~= +1.14~$\pm$~0.08. 
It is also in reasonable accord with the Bi\'emont \& Lowe 
(1993)\nocite{bie93} value, \eps{Dy}$_\odot$~= +1.20~$\pm$~0.06.

Synthetic spectra of 24--35 lines were used in the Dy abundance derivations 
for the \rpro-rich low-metallicity giants (Table~\ref{tab7}).
The analyses were straightforward, as many \ion{Dy}{2} lines in each 
star's spectrum were strong and unblended.
This led to very well-determined mean abundances (Tables~\ref{tab4}
and \ref{tab5}).

\subsection{Thulium\label{tmtext}}

Tm ($Z$~=~69) has one naturally-occurring isotope, \iso{169}{Tm}.
This element is one of the least abundant of the REs:
\eps{Tm}$_{met}$~= 0.11~$\pm$~0.06 (Lodders 2003)\nocite{lod03}.
Therefore \ion{Tm}{2} transitions in solar and stellar spectra are
weak, and relatively few can be employed in abundance analyses.
Moore \etal\ (1966)\nocite{moo66} list only 10 \ion{Tm}{2}
identifications in their solar line compendium; all of these
lie at wavelengths $\lambda$~$<$~4300~\AA.

We considered the 146 \ion{Tm}{2} lines investigated by 
Wickliffe \& Lawler (1997)\nocite{wic97}.
That study reported laboratory experimental transition probabilities 
derived from their branching fractions and the radiative 
lifetimes of Anderson, Den Hartog, and Lawler (1996)\nocite{and96}.
The relative strengths of these lines are displayed in panel (c) of 
Figure~\ref{f1}.
Inspection of this plot suggests that few detectable \ion{Tm}{2} lines will
be found redward of 4000~\AA, in accord with the Moore \etal\ 
(1966)\nocite{moo66} identifications.

As in the case of Pr (\S\ref{prtext}), we began our search for suitable 
\ion{Tm}{2} transitions with \csthree, since they should stand out most 
clearly among the weaker Fe-peak contaminants in this star's spectrum.
Only nine lines were sufficiently strong and unblended to warrant 
further investigation.
We computed synthetic spectra for each of these candidate features.
Although Tm is an odd-$Z$, odd-$A$ atom with a non-zero nuclear
spin ($I$~=~$\frac{1}{2}$), inspection of the chosen \ion{Tm}{2} lines
in very high-resolution NSO FTS spectra showed that hyperfine splitting
is very small, and could be safely ignored in the calculations.

\begin{figure}
\epsscale{0.9}
\plotone{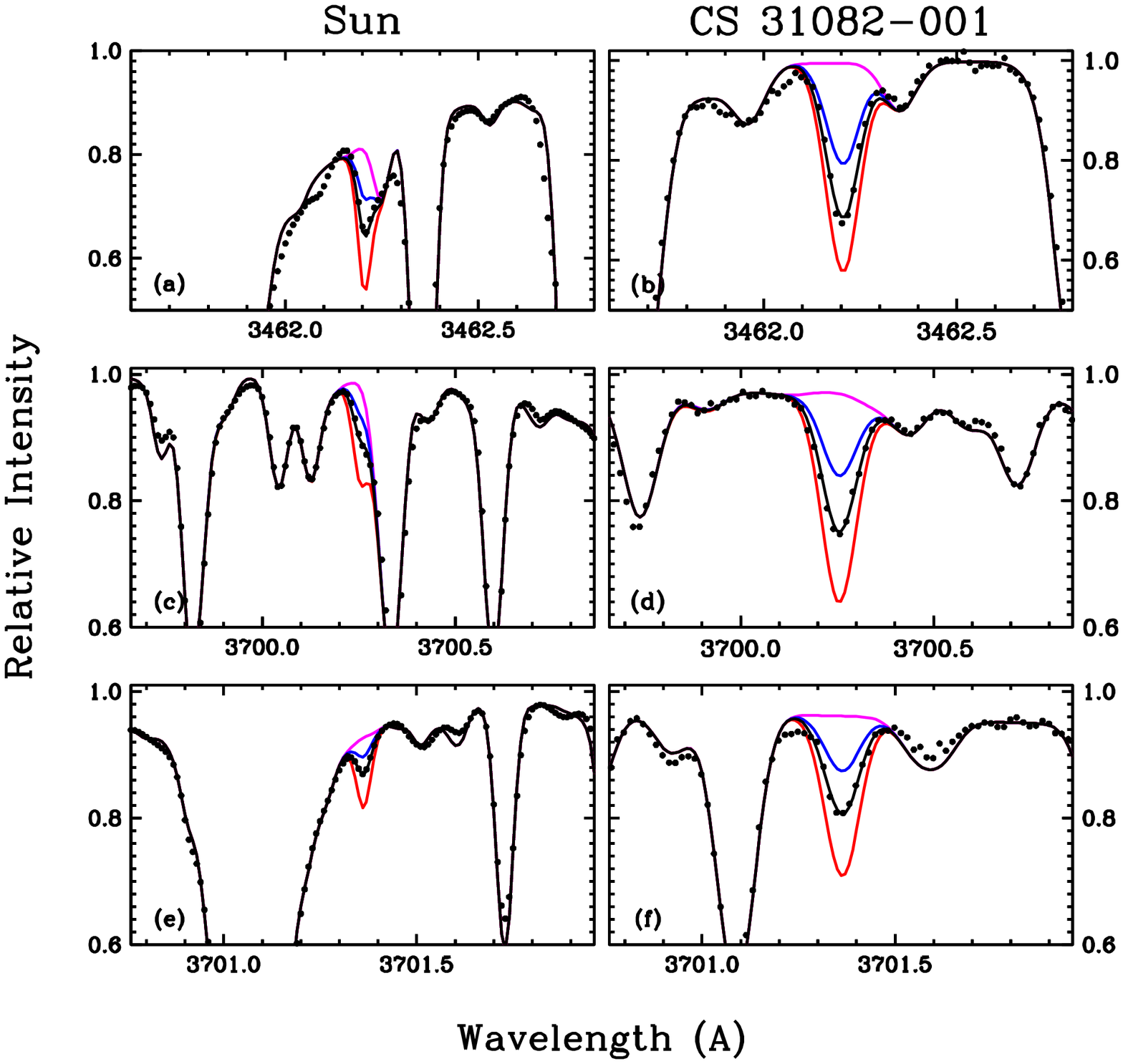}
\caption{
\footnotesize
Observed and synthetic spectra of the Sun (left-hand
panels (a), (c), (e)) and CS~31082-001 (right-hand panels
(b), (d), (f)) for the three \ion{Tm}{2} lines that contribute
to the solar abundance estimate.
In each panel, the points represent the observed spectrum.
The magenta line is the spectrum computed with no contribution
from \ion{Tm}{2}; the black line is the best-fitting synthesis (with the
Tm abundance given in Table~\ref{tab8}); and the red and blue lines
are the syntheses computed with Tm abundances altered by $\pm$0.3~dex
from the best value.
The solar spectrum is that of Delbouille \etal\ (1973), but sampled at
a wavelength step size of 0.01~\AA\ for display purposes.
\label{f6}
}
\end{figure}

Our synthetic spectra of \ion{Tm}{2} lines for the solar photosphere
showed that only three of them could be used for abundance analysis.
The synthetic/observed spectrum matches for these lines in the
solar photosphere are displayed in Figure~\ref{f6},
along with those for \csthree.
It is clear that each of these lines is weak and blended in the 
photospheric spectrum, while being much stronger and cleaner
in the \rpro-rich low metallicity giant star.

These lines and their photospheric abundances are listed in 
Table~\ref{tab8}. 
We derive a formal mean abundance (Table~\ref{tab3}) of
\eps{Tm}$_\odot$~= +0.14~$\pm$~0.02 ($\sigma$~= 0.04).
Caution obviously is warranted here.
Probably the $\sigma$ value is a truer estimate of the abundance
uncertainty than the standard deviation of the mean.
However, this photospheric abundance is in reasonable agreement with
the meteoritic value, \eps{Tm}$_{met}$~= +0.11~$\pm$~0.06.

More \ion{Tm}{2} features could be employed in the abundance 
determinations for the \rpro-rich low-metallicity giants 
(Table~\ref{tab8}). 
Their mean values (Tables~\ref{tab4} and \ref{tab5})
were based on 5--7 lines per star.
For stars analyzed previously by our group, the new Tm abundances agree
with the published values to within the uncertainty estimates.
The Tm abundance for \csthree will be discussed along with this star's
other REs in \S\ref{rrichstars}.

\subsection{Ytterbium\label{ybtext}}

Yb ($Z$~=~70) has seven naturally-occurring isotopes, six of which
are major components of its solar-system abundance:
\iso{168}{Yb},  $\ll$1\%;
\iso{170}{Yb},  3.04\%;
\iso{171}{Yb},  14.28\%;
\iso{172}{Yb},  21.83\%;
\iso{173}{Yb},  16.13\%;
\iso{174}{Yb},  31.83\%; and
\iso{176}{Yb},  12.76\% (Lodders 2003\nocite{lod03}).
The atomic structure of \ion{Yb}{2} is similar to that of \ion{Ba}{2},
with a $^2$S ground state and first excited state more than 2.5~eV above
the ground state.
Therefore this species has very strong resonance lines at 3289.4 and
3694.2~\AA\ as the only obvious spectral signatures of this element.
All other \ion{Yb}{2} lines are expected to be extremely weak.

The \ion{Yb}{2} resonance lines have complex hyperfine and isotopic
substructures that broaden their absorption profiles by 0.06~\AA\
and must be included in synthetic spectrum computations.
In the Appendix we discuss the literature sources for the 
resonance lines and tabulate their substructures in a form useful for
stellar spectroscopists.
Moore \etal\ (1966)\nocite{moo66} identified major \ion{Fe}{1}, 
\ion{Fe}{2}, and \ion{V}{2} contaminants to the 3289.4~\AA\ line,
and our synthetic spectra confirmed that \ion{Yb}{2} is a small 
contributor to the total feature.
From our synthetic spectra of the 3694.2~\AA\ line we derived 
\eps{Yb}$_\odot$~= +0.86~$\pm$~0.10 (Table~\ref{tab3}), in 
reasonable agreement with \eps{Yb}$_{met}$~= +0.94~$\pm$~0.03.
The large uncertainty attached to our photospheric abundance arises
from a variety of sources: (a) reliance on a single \ion{Yb}{2}
line; (b) its large absorption strength, which increases the dependence
on adopted microturbulent velocity; (c) the contaminating presence of
the strong \ion{Fe}{1} 3694.0~\AA\ line; and (b) closeness of this
spectral region to the Balmer discontinuity.

We then synthesized the 3289 and 3694~\AA\ lines in the stellar spectra.
However, these are \rpro-rich stars, and the isotopic mix in a pure
\rpro\ nucleosynthetic mix is different than that of the solar system
(\rpro\ and \spro) combination.
For our computations we adopted (see Sneden \etal\ 2008)\nocite{sne08}:
\iso{168,170}{Yb},  0.0\%;
\iso{171}{Yb},     17.8\%;
\iso{172}{Yb},     22.1\%;
\iso{173}{Yb},     19.0\%;
\iso{174}{Yb},     22.7\%; and
\iso{176}{Yb},     18.4\%. 

The Yb contribution to the 3289~\AA\ feature is very large in the 
\rpro-rich stars.
In the most favorable case, \csthree, Yb accounts for roughly 75\% of the
total blend.
Unfortunately, the contributions of the contaminants (mostly \ion{V}{2})
cannot be assessed accurately enough for this line to be a reliable
Yb abundance indicator.
The synthetic/observed spectral matches of the 3694~\AA\ line 
provide the new Yb abundances listed in Tables~\ref{tab4} 
and \ref{tab5}.
These values are consistent with the ones reported in the
original papers on these stars.
However, while the \ion{Yb}{2} absorption dominates
that of the possible metal-line contaminants, the Balmer lines in this
spectral region are substantially stronger in these low-pressure giant
stars than they are in the solar photospheric spectrum.
In particular, \ion{H}{1} lines at 3691.6 and 3697.2~\AA\ significantly
depress the local continuum at the \ion{Yb}{2} wavelength.
Caution is warranted in the interpretation of these Yb abundances.

\subsection{Lutetium\label{lutext}}

Lu ($Z$~=~71), has two naturally-occurring isotopes: 
\iso{175}{Lu}, 97.416\%; and 
\iso{176}{Lu}, 2.584\% (Lodders 2003\nocite{lod03}).
It is the least abundant RE: \eps{Lu}$_{met}$~= 0.09~$\pm$~0.06 
(Lodders\nocite{lod03}).
\ion{Lu}{2} has a relatively simple structure, with a $^1$S ground state.
It has no other very low-energy states; the first excited level lies
1.5~eV above the ground state.
This ion with only two valence electrons has relatively few strong lines
in the visible and near UV connected to low E.P. levels, although most
of the prominent lines have well-determined experimental transition
probabilities.

We considered only the \ion{Lu}{2} transitions of Quinet \etal\ 
(1999)\nocite{qui99}, using their experimental branching fractions and 
lifetime measurements by Fedchak \etal\ (2000)\nocite{fed00} to 
determine \ion{Lu}{2} transition probabilities.
These are are listed, along with wavelengths and excitation energies, 
in Table~12 of Lawler \etal\ (2009)\nocite{law09}.
The combination of a small solar-system Lu abundance and the (unfavorable) 
atomic parameters produces very small relative strength factors for these 
lines, as shown in panel (d) of Figure~\ref{f1}.
No line even rises to our defined ``weak-line'' threshold of usefulness.
Moore \etal\ (1966)\nocite{moo66} lists only 3077.6, 3397.1, and
3472.5~\AA\ \ion{Lu}{2} identifications in their solar line compendium, and
all of these lines appear to be blended.

We made a fresh search for detectable lines of \ion{Lu}{2}, and 
succeeded mainly in confirming the results of a previous investigation by 
Bord, Cowley, \& Mirijanian (1998)\nocite{bor98}.
Those authors argued that all of the lines identified by Moore \etal\
(1966)\nocite{moo66} are unsuitable for solar Lu abundance work.
They quickly dismissed the 3077.6 and 3472.5~\AA\ lines and performed 
an extended analysis of 3397.1~\AA.
Synthetic spectrum computations around this feature (see their
Figures~2 and 3) convinced them that molecular NH dominates the 
absorption at the \ion{Lu}{2} wavelength.
Our own trials produced the same outcome.

Bord \etal\ (1998)\nocite{bor98} detected \ion{Lu}{2} 6221.9~\AA\ in
the Delbouille \etal\ (1973) photospheric spectrum.
This line is extremely weak, EW~$\sim$ 1~m\AA, and its hyperfine 
substructure spreads the absorption over $\sim$0.5~\AA.
The complex absorption profile of this line (see their Figure~4) 
actually increases one's confidence in its identification in the
photospheric spectrum.
Bord \etal\ reported \eps{Lu}$_\odot$ =~+0.06 with an estimated $\pm$0.10 
uncertainty from this line.

We repeated their analysis, using the hyperfine substructure pattern
given in Table~13 of Lawler \etal\ (2009)\nocite{law09}, and derived 
\eps{Lu}$_\odot$~= +0.12~$\pm$~0.08 (Table~\ref{tab3}), where the 
error reflects uncertainties in matching synthetic and observed 
feature profiles.
This photospheric abundance is consistent with the Bord \etal\ 
(1998)\nocite{bor98} value and with the meteoritic abundance quoted
above, given the uncertainties attached to each of these estimates.
Our lack of success in identifying other Lu abundance indicators in
the solar photospheric spectrum suggests that prospects are poor for
reducing its error bar substantially in the future.

We also attempted to study the 3397 and 6621~\AA\ lines
in our sample of \rpro-rich low metallicity giants.
Absorption by \ion{Lu}{2} at 3397.1~\AA\ is certainly present in the
spectra of at least \cstwo\ and \csthree.
Unfortunately, the lower resolutions of our stellar spectra compared to
that of the solar spectrum creates more severe blending of the Lu
transition with neighboring lines, and NH contamination of the total
feature still creates substantial abundance ambiguities.
The 6221.9~\AA\ line should be present, albeit very weak, in these
stars.
However, our spectra (when they extend to this wavelength range) lack
the S/N to allow meaningful detections.
We therefore cannot report Lu abundances for these \rpro-rich stars.

\section{RARE EARTH ABUNDANCE DISTRIBUTIONS IN THE SUN AND 
R-PROCESS-RICH STARS\label{rareab}}

\subsection{The Sun and Solar System\label{solarab}}

With new analyses of Pr, Dy, Tm, Yb, and Lu we now have determined 
abundances for the entire suite of REs in the solar photosphere.
In Table~\ref{tab3} we merge the results of this and our
previous papers.
Missing from the list is of course Pm (Z~=~61), whose longest-lived
isotope, \iso{145}{Pm}, is only 17.7~years  
(Magill, Pfennig, and Galy 2006\nocite{mag06}).
We also chose not to include a photospheric value for Ba,
whose few transitions are so strong that their solar absorptions 
cannot be reliably modeled in the sort of standard photospheric 
abundance analysis that we have performed.

The photospheric abundance uncertainties quoted in Table~\ref{tab3}
are combinations of internal ``scatter'' factors (mainly continuum 
placement, observed/synthetic matching, and line blending problems) and 
external ``scale'' factors (predominantly solar model atmosphere
choices).
These issues are discussed Lawler \etal\ (2009)\nocite{law09} and
in previous papers of this series.
We remind the reader that our abundance computations have been 
performed with the traditional assumptions of LTE and 1D
static atmosphere geometry.
Very little has been done to date to explore the effects of these 
computational limitations for RE species in the solar atmosphere.
Mashonkina \& Gehren (2000)\nocite{mas00} have performed non-LTE 
abundance analyses of Ba and Eu, but their photospheric
abundances are not substantially different from LTE results.
There have been efforts to model the solar spectrum with more realistic 
3-dimensional (3D) hydrodynamic models; see the summary in Grevesse,
Asplund, \& Sauval (2007)\nocite{gre07}, and references therein.
These studies so far have reported new solar abundances 
only for the lighter elements (CNO, Na$-$Ca, and Fe).
Generally the 3D non-LTE line modeling efforts yield
lower abundances: comparing the photospheric values in Grevesse 
\etal\ to those of the older standard compilation of Anders \&
Grevesse (1989)\nocite{and89}, 
$<\delta$log~$\epsilon>$~= $-$0.12~$\pm$~0.03 ($\sigma$~=~0.09, for
11 elements that can be studied with photospheric spectra).
We thus expect that any RE abundance shifts with 3D modeling would be
similar from element to element, leaving their abundance ratios 
essentially unchanged.
Future studies to explore these effects in detail will be welcome.

\begin{figure}
\epsscale{0.8}
\plotone{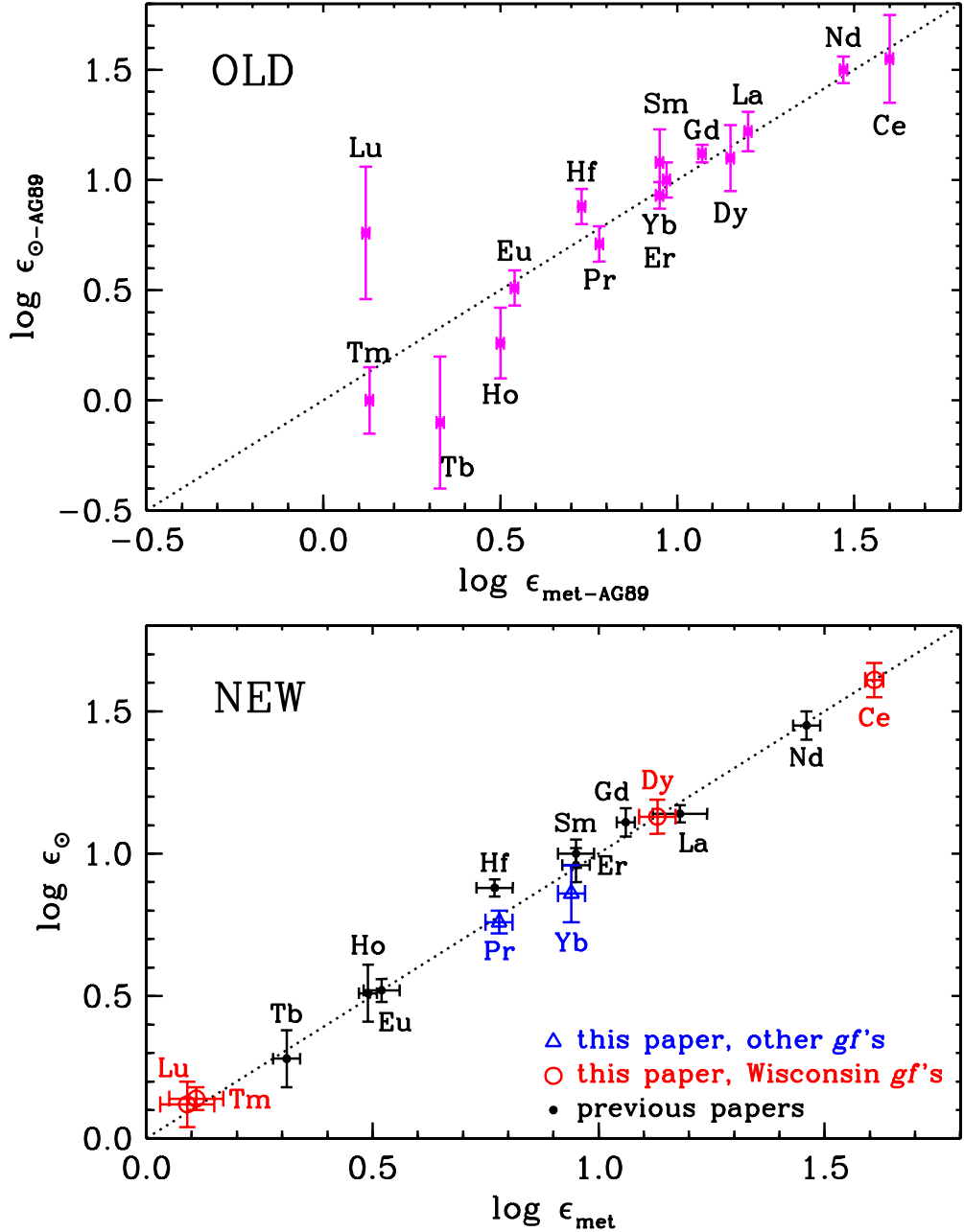}
\caption{Comparisons of solar-system meteoritic and solar photospheric 
abundances of the RE elements.
In the top panel, the ``OLD'' abundances are the recommended values from
two decades ago (Anders \& Grevesse 1989).
In the bottom panel, the ``NEW'' meteoritic values are from (Lodders 
2003), and the solar photospheric abundances are from in this study 
and previous papers of this series.
We separate the new photospheric results into three groups, using red open
circles to denote those elements whose abundances are based on transition
probabilities published by the Wisconsin group, blue open triangles
for those elements whose abundances are based on other sources for
the transition probabilities, and black dots for abundances determined 
in earlier papers of this series.
\label{f7} \footnotesize
}
\end{figure}

In Figure~\ref{f7} we compare RE photospheric abundances to
their meteoritic values.
In the top panel the ``OLD'' values are best estimates by Anders \& 
Grevesse (1989\nocite{and89}).
While the average agreement is good, significant discrepancies between 
individual abundances are evident, particularly at the low-abundance end.
Formally, a simple mean is 
$<\log \epsilon_{\odot-AG89} - \log \epsilon_{met-AG89}>$~= 
0.00~$\pm$~0.06 ($\sigma$~=~0.22).
In the bottom panel, the ``NEW'' meteoritic abundances 
(Lodders 2003)\nocite{lod03} are correlated with our ``NEW''
photospheric ones (Table~\ref{tab3}).
The data sources are denoted by different symbols in the figure:
red open circles for photospheric abundances newly determined here
and in Lawler \etal\ (2009)\nocite{law09} for which Wisconsin-group 
lab data have been used;
black filled circles for abundances reported in our previous papers;
and blue open triangles for two elements with transition probability
data adopted from other literature sources.
Clearly the agreement is excellent: for 15 elements the formal
mean difference is 
$<\log \epsilon_\odot - \log \epsilon_{met}>$~= 0.01~$\pm$~0.01
($\sigma$~=~0.05).
No trends are discernible with the source of atomic data, or the abundance 
levels (as shown in the figure), or the number of lines that contribute 
to the photospheric abundances (Table~\ref{tab3}).
With the possible exception of Hf (discussed in Lawler \etal\ 
2007\nocite{law07} and in \S\ref{discuss}), and with repeated
cautions about the photospheric abundances deduced from only one
or two transitions, the two primary sources of primordial Solar-System 
abundances appear to be in complete accord.

\subsection{The \rpro-Rich Low Metallicity Giant Stars\label{rrichstars}}

Rare-earth abundances for the five \rpro-rich stars from this
and our previous papers are collected in Tables~\ref{tab4} and
\ref{tab5}.
For all stars the Pr, Dy, Tm, and Yb abundances are, of course,
newly determined in this paper.
We chose also to redo all the Ba abundances via new synthetic 
spectrum calculations, to ensure that these were determined in
a consistent manner.
We also performed new analyses for selected elements in individual 
stars (\eg, Tb in \hdone) when the original papers either did not report 
abundance values or did so with now-outdated atomic data.

Of particular interest is the very \ncap-enhanced star \csthree, 
which is a recent addition to our \rpro-rich star list.
This star gained notoriety as the first \rpro-rich star with
a convincing detection of U, a long-lived radioactive element of
great interest to cosmochronology (Cayrel \etal\ 2001\nocite{cay01}).
The first and most complete study of this star was published by
Hill \etal\ (2002)\nocite{hil02}.
The mean difference between our RE abundances for this star 
and theirs is 
$<\log \epsilon_{Hill} - \log \epsilon_{us}>$~= $-$0.05~$\pm$~0.03
($\sigma$~=~0.10, 12 elements in common).
We also compared our \csthree\ abundances with those of Honda \etal\
(2004\nocite{hon04}), with similar results:
$<\log \epsilon_{Honda} - \log \epsilon_{us}>$~= +0.07~$\pm$~0.03
($\sigma$~=~0.09, 12 elements in common).
The mean offsets are very small, and reflect minor differences
in model atmospheres, observed spectra, analytical techniques,
and atomic data choices.
The element-to-element scatters are also reasonable, given the
use of many more transitions in our study (a total of 342, 
Table~\ref{tab4}) compared to 95 in Hill \etal\ and
49 in Honda \etal\
Note that some portion of the $\sigma$'s in these comparisons
arises because the Tb abundance differences are offset by $\simeq$0.2~dex
from the mean differences (we derive larger values).
Investigation of this one anomaly is beyond the scope of this work.

The abundance standard deviations of samples ($\sigma$) and of means that
are given in Tables~\ref{tab4} and \ref{tab5} refer
to internal (measurement scatter) errors only.
To investigate scale uncertainties, we determined the abundance sensitivities
of eight RE elements to changes in model parameters
(\teff, \logg, [M/H], \vmicro), to changes in the adopted model
atmosphere grid, and to changes to line computations to better
account for continuum scattering opacities.
In Table~\ref{tab9} we summarize the results of these exercises.
We began with a ``baseline'' model atmosphere from the 
Kurucz (1998)\nocite{kur98} grid with parameters
\teff~=~4750~K, \logg~=~1.5, [M/H]~=~--2.5, and \vmicro~=~2.0.
Such a model is similar to the ones adopted for the \rpro-rich
giants (Table~\ref{tab2}).
We derived abundances with this model for 1-4 typical transitions each 
of the elements for the program star \csthree.
Full account was taken of hyperfine and isotopic substructure for La, Pr, 
Eu, and Yb.
We then repeated the abundance derivations for models with parameters
varied as indicated in Table~\ref{tab9}, including a trial
using a model with baseline parameters taken from the new MARCS
(Gustafsson \etal\ 2008\nocite{gus08}) grid.\footnote{
Available at http://marcs.astro.uu.se/}
The inclusion of scattering in computations of continuum source
functions, a new feature in our analysis code,
is described in Sobeck \etal\ (2008)\nocite{sob09}

The Table~\ref{tab9} quantities are differences between
abundances of the individual models and those of the baseline model.
The uncertainties in stellar model parameters given in the original 
\rpro-star papers are typically $\pm$150~K in \teff, $\pm$0.3 in \logg,
$\pm$0.2~\kmsec\ in \vmicro, and $\pm$0.2 in [M/H] metallicity.
Application of these uncertainties to the model parameter dependences
of Table~\ref{tab9} suggests that [M/H] and \vmicro\ choices are not
important abundance error factors.
Temperature and gravity values obviously play larger roles.  
However, while the absolute abundances of individual elements 
change with different \teff\ and \logg\ choices, the relative abundances
generally do not; in most cases, all RE abundances move in lock step.
Assuming here that the atmosphere parameter uncertainties are 
uncorrelated, we estimate total abundance uncertainties for each 
RE element to be $\sim$0.15$-$0.20, but the abundance ratios
have uncertainties of $\sim$0.01$-$0.05 (the exception is Yb,
represented by only one very strong line in the UV spectral region;
see \S\ref{ybtext}).
More detailed computations that consider departures from LTE among
RE first ions in the atmospheres of very metal-poor giant stars
should be undertaken in the future.
Some first steps in this direction have been undertaken for Ba and Eu by
Mashonkina \etal\ (2008)\nocite{mas08}, but such calculations will need
to be repeated for many REs to understand the magnitude of corrections
to the abundances reported here.

\section{DISCUSSION\label{discuss}}

\begin{figure}
\epsscale{1.0}
\plotone{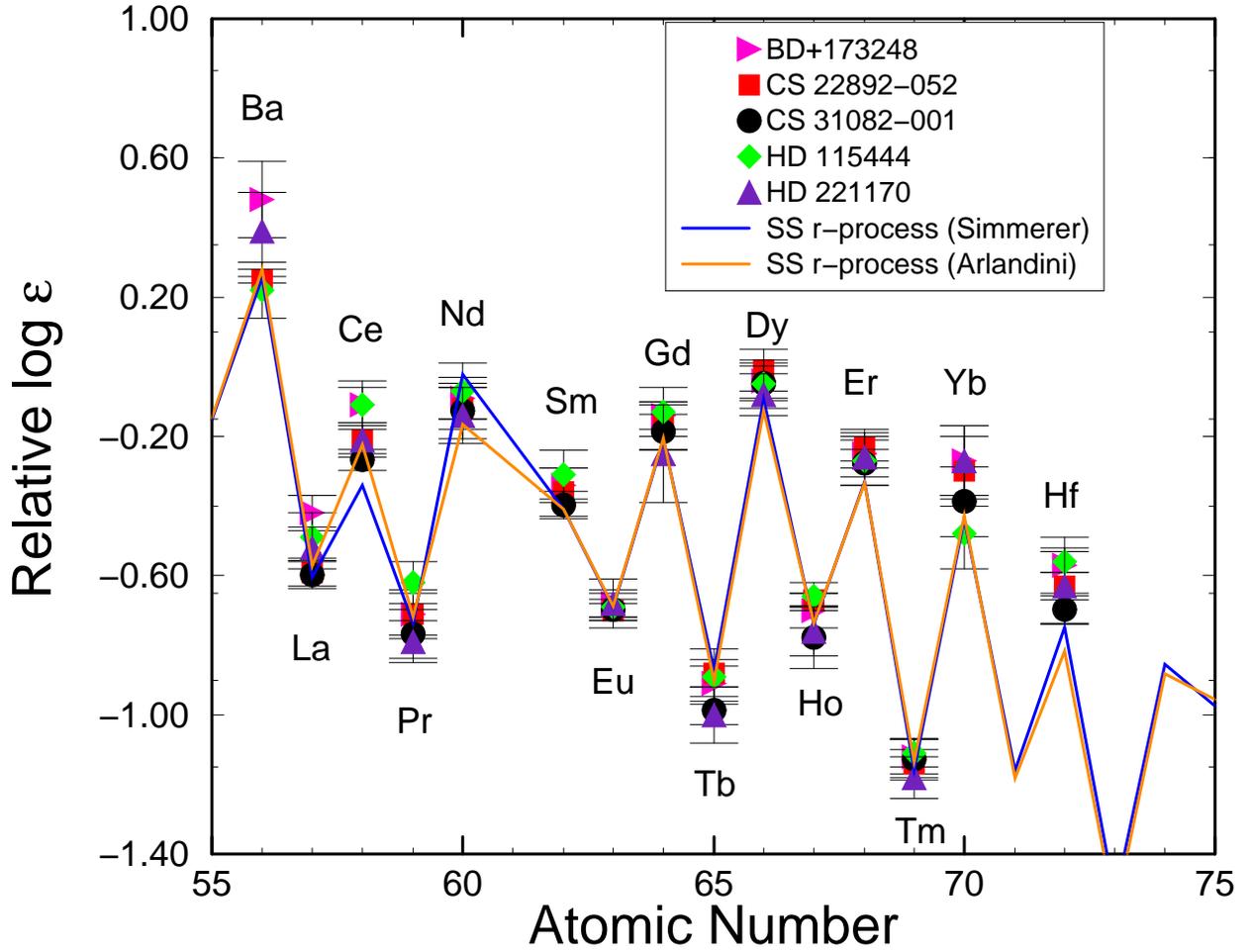}
\caption{Comparison of the newly derived RE abundances in five r-rich halo 
stars with predictions for solar system r-process only abundances from 
Arlandini \etal\ (1999) and Simmerer \etal\ (2004). 
For each star the abundances have been normalized at Eu.
\label{f8} \footnotesize
}
\end{figure}

\begin{figure}
\epsscale{1.0}
\plotone{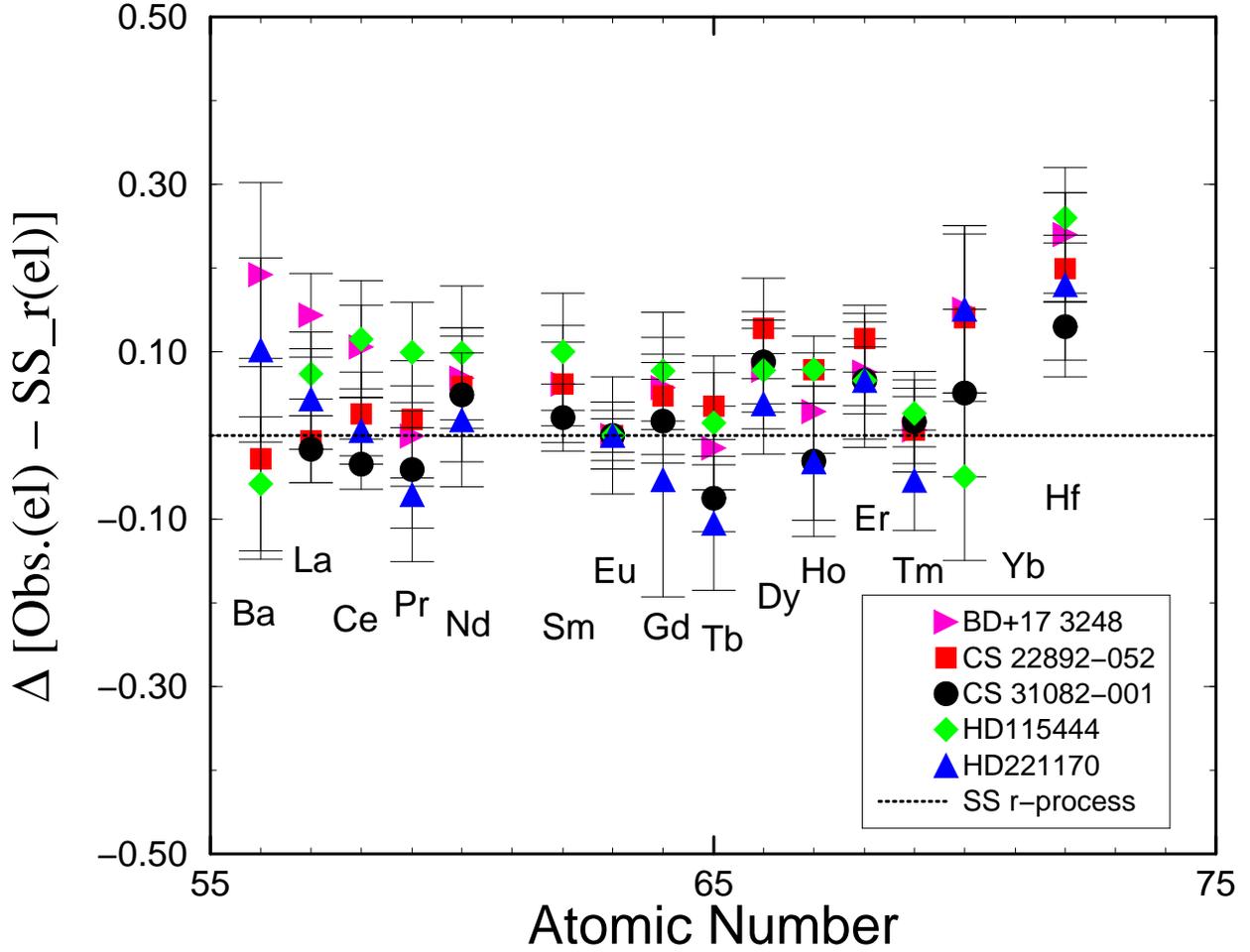}
\caption{Comparison of the newly derived RE abundances
in five r-rich halo stars to  
the solar system r-process only value from
Arlandini \etal\ (1999). 
For each star the abundances have been normalized at Eu.
The dotted line indicates a perfect agreement between the stellar and 
solar system r-only values. The error bars are the sigma values
listed for each star in Tables~\ref{tab4} and \ref{tab5}. 
\label{f9} \footnotesize
}
\end{figure}

We illustrate the RE abundances for \bd, \cstwo, \csthree, 
\hdone\ and \hdtwo\ in Figures~\ref{f8} and \ref{f9}.
For each star the abundances have been normalized at Eu, a 
predominantly $r$-process element.
In Figure~\ref{f8} these relative abundances are shown in comparison 
to the Solar System \rpro-only predictions from Arlandini \etal\ 
(1999)\nocite{arl99} and Simmerer \etal\ (2004)\nocite{sim04}.
We note first the excellent star-to-star (relative abundance) 
agreement. 
Early RE abundance distributions of \ncap-rich metal-poor stars 
indicated large star-to-star scatter for a number of individual elements 
(\eg, Luck \& Bond 1985\nocite{luc85}, Gilroy \etal\ 1988\nocite{gil88}). 
The combination of substantially better S/N and resolution of
the stellar spectra and the experimental initiatives of this series of 
papers has dramatically reduced that scatter -- all the RE elements are 
now in very good (relative) agreement for these five halo stars.

Figure~\ref{f8} also uses solid lines to illustrate the solar-system 
\rpro-only meteoritic abundances determined by Simmerer \etal\ 
(2004)\nocite{sim04} and Arlandini \etal\ (1999)\nocite{arl99}.
In both cases, these values were computed by subtracting the \spro-only
abundances from the total elemental abundances.
The ``classical'' method (Simmerer \etal) matches smooth $\sigma$N$_s$ 
curves to those isotopes of \ncap\ elements whose production is
essentially all due to the \spro, and infers from those empirical 
curves the \spro\ amounts of elements that can be produced by both 
the \rpro\ and \spro.
The solar system \rpro\ abundances are then just the residuals 
between total elemental and \spro\ amounts.
The ``stellar'' method (Arlandini \etal) uses theoretical 
models of \spro\ nucleosynthesis instead of empirical \spro\ 
abundance curves, and again infers the \rpro\ amounts by subtraction.

Our stellar abundances compare very well with the relative 
solar system \rpro\ distributions.
In the past we and other investigators have found overall agreement,  
but on a more approximate scale. 
The new abundance determinations shown in Figure~\ref{f8} tighten
the comparison, with deviations between the stellar and 
solar system r-process curves of typically less than 0.1 dex --
probably the practical limit of what is currently possible.
These abundance comparisons strongly support many other studies 
(see Sneden \etal\ 2008\nocite{sne08}, and references therein) arguing 
that essentially the same process was responsible for the formation of 
all of the \rpro\ contributions to these elements early in the history 
of the Galaxy in the element progenitor stars to the presently-observed
\rpro-rich halo stars. 

Despite this general level of elemental abundance consistency, there are 
some interesting deviations. 
In particular, the two solar system \rpro\ predictions differ by about
0.1~dex for the elements Ce and Nd (Table~\ref{tab3}).
In both cases the stellar model predictions from Arlandini \etal\ 
(1999)\nocite{arl99} give a better fit to the stellar abundance data 
than do the standard model predictions from Simmerer \etal\ (2004). 
This suggests that the Arlandini \etal\ \rpro\ distribution might be
superior for such abundance comparisons.
This has been noted previously by others (\eg, Roederer \etal\ 
2008\nocite{roe08}) for isotopic studies.

There is also still some star-to-star scatter particularly at Ba, 
with several stellar elemental abundances appearing somewhat higher 
than the solar system r-process curves. 
This can be seen more clearly in Figure~\ref{f9}, where we 
illustrate the difference between the relative (scaled to Eu) stellar 
RE and the scaled solar system r-process abundances (Arlandini \etal\ 
1999\nocite{arl99}) in the five \rpro-rich stars. 
While most of the individual elemental abundance data lie 
close to the dotted line (indicating perfect agreement with the solar
\rpro), Ba and Yb have significant star-to-star scatter.
But both elements have inherent observational problems, as they are
represented by only a few very strong transitions that have multiple 
isotopic components whose relative abundances are sensitive to the 
relative $r$-/\spro\ dominance (recall the Yb discussion in \ref{ybtext}). 
Abundance determinations for Yb and Ba are less reliable than those
of most other RE elements, and should be treated with caution.

We also note that for \bd\ the RE abundances relative to Eu 
appear to be somewhat higher their values in the other stars, 
particularly for the predominantly \spro\ elements Ba and La. 
\bd\ has a metallicity of [Fe/H]~$\simeq$ --2.1 (Cowan \etal\ 
2002\nocite{cow02}), so this star might be showing the signs of the 
onset of Galactic $s$-processing, which occurs at approximately that
metallicity (Burris \etal\ 2000\nocite{bur00}). 
On the other hand \hdtwo\ with a similar metallicity (Ivans \etal\ 
2006\nocite{iva06}) does not seem to show the same deviations for the 
\spro\ elements, and thus the deviations for \bd\ may be 
specific to that star. 

\begin{figure}
\epsscale{1.0}
\plotone{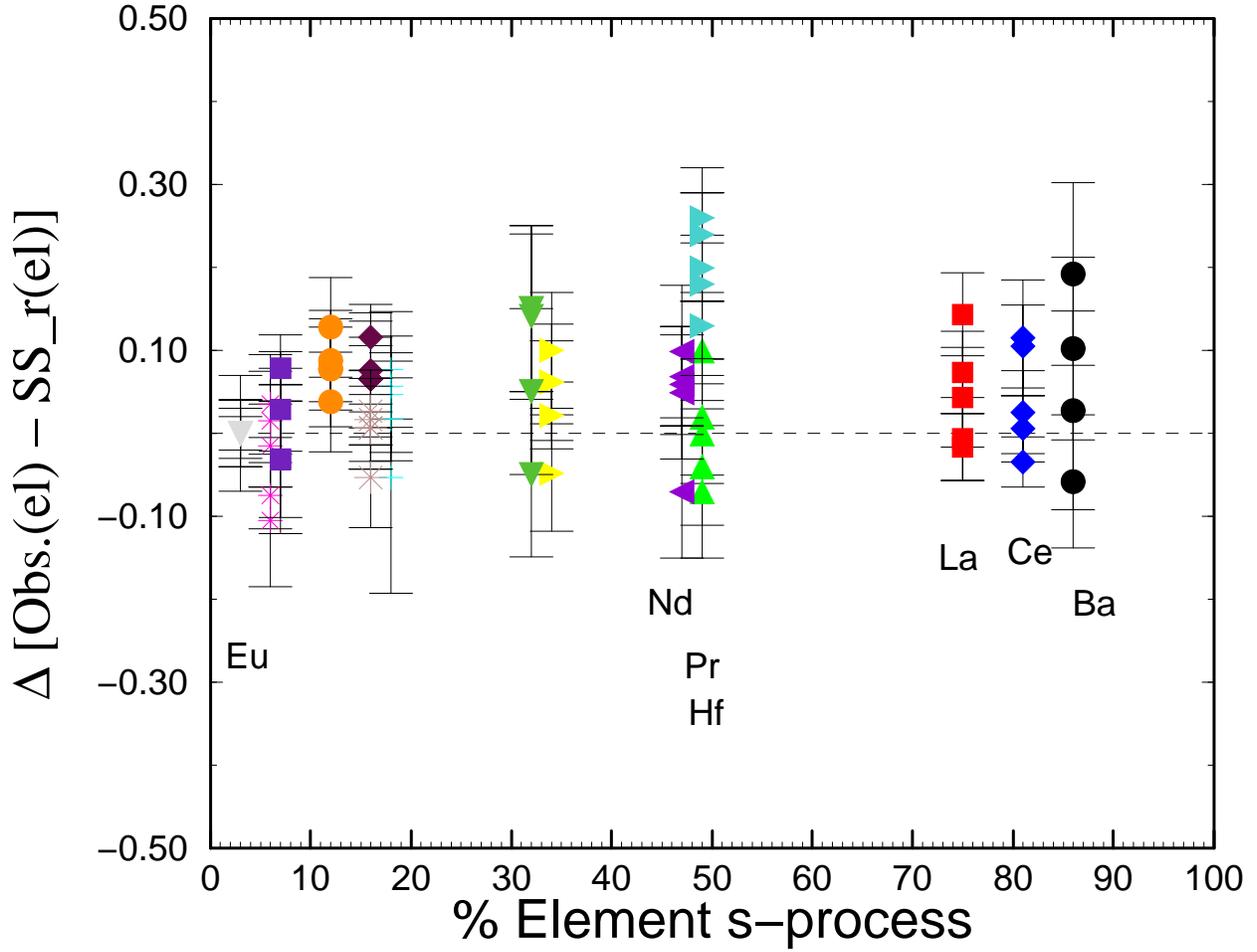}
\caption{
Comparison of the newly derived RE abundances
in five r-rich halo stars to
the solar system r-process only value from
Arlandini \etal\ (1999) as a function of percentage of
the solar system elemental s-process.
The dashed line indicates a perfect agreement between the stellar and
solar system r-only values.
For clarity in this figure, a different color has been used for each element.
\label{f10} \footnotesize
}
\end{figure}

We examine whether there is any correlation between the deviation of 
the stellar abundances from the solar system r-process values and the
\spro\ percentage of those elements in solar system material 
(from Simmerer \etal\ 2004)\nocite{sim04} in Figure~\ref{f10}. 
It is clear that there is little if any secular trend with the abundance 
differences with increasing solar-system \spro\ abundance percentage.
This lack of correlation was also found specifically for the element 
Ce by Lawler \etal\ (2009).\nocite{law09} 

\begin{figure}
\epsscale{1.0}
\plotone{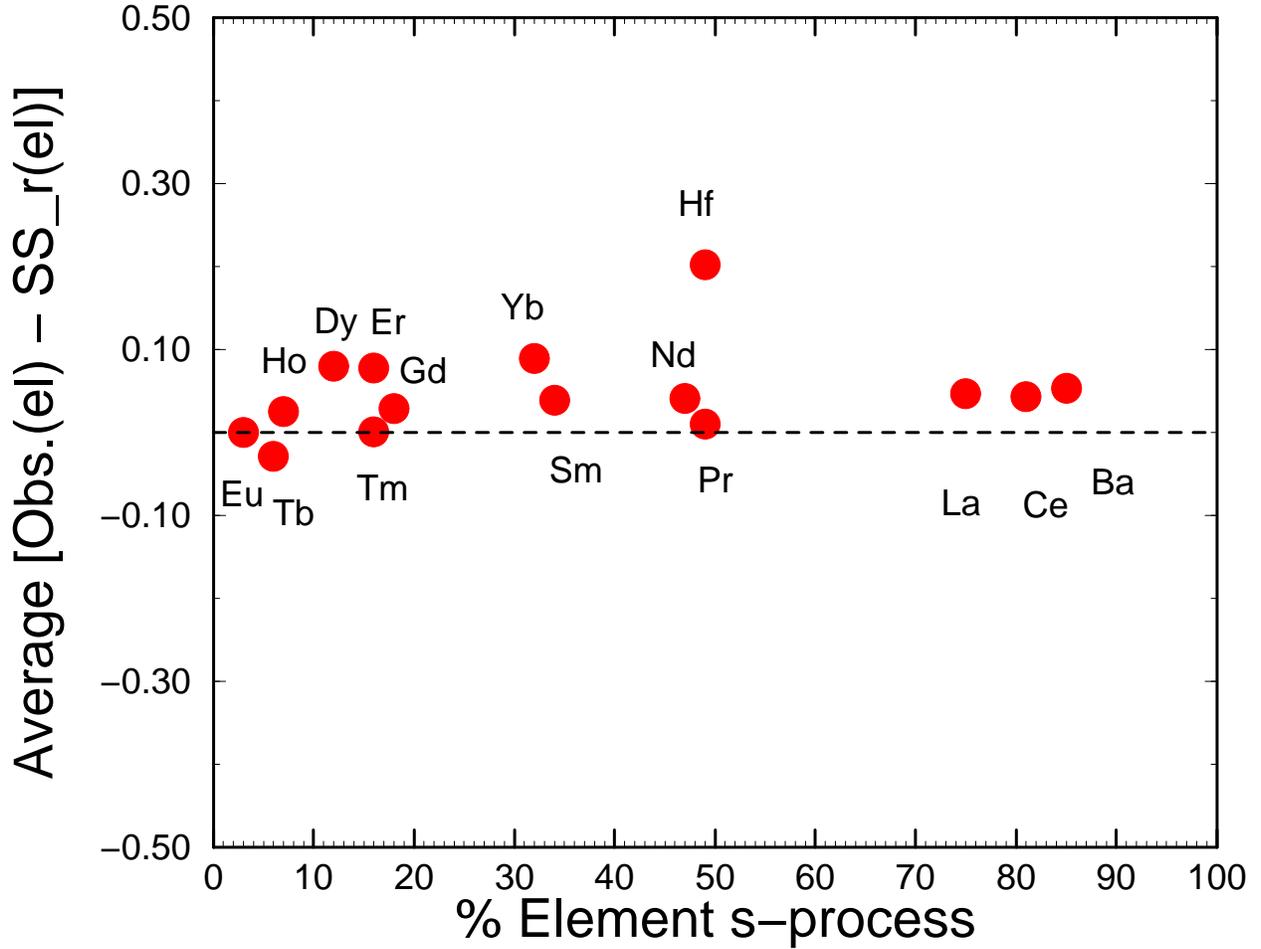}
\caption{
Averages of the stellar elemental abundance offsets of the five stars with 
respect to the solar system r-process only value from Arlandini \etal\ (1999) 
as a function of percentage of the solar system elemental s-process.
For each star included in the average the abundance offsets have been 
normalized at Eu.
The dotted line indicates a perfect agreement between the average stellar and
solar system r-only values. 
\label{f11} \footnotesize
}
\end{figure}

To get a clearer sense of the overall abundance agreement with the 
solar-system \rpro\ abundances, we show in Figure~\ref{f11} 
the arithmetic averages of the elemental abundance offsets 
(from Figure~\ref{f10}) for the five stars, again as a function 
of \spro\ percentage.
Obviously these average offsets with respect to the solar-system
\rpro\ values are very small. 
Including all elements the mean of the average offsets is 
$\log \epsilon$ = 0.05 ($\sigma$ = 0.05). 
Previously Lawler \etal\ (2007)\nocite{law07} had found that 
the observed average stellar abundance ratio of Hf/Eu in a group of
metal-poor halo stars is larger than previous estimates of the 
solar-system \rpro-only value, suggesting a somewhat larger contribution
from the \rpro\ to the production of Hf.  
Our new analysis supports that finding, as the average Hf offset is larger
than all of the other elemental abundances. 
If the solar system \rpro\ contribution was larger it would drive 
down the average offset illustrated in Figure~\ref{f11}. 
Ignoring the Hf results, the mean of the average offsets for all of 
the other RE elements is 0.04 ($\sigma$ = 0.03). 
This is essentially a perfect agreement within the limits of our 
observational and experimental uncertainties, as well as the uncertainties 
(observational and theoretical) associated with the solar system 
\rpro-only abundance values.

\section{CONCLUSIONS\label{conclude}}

We have determined new abundances of Pr, Dy, Tm, Yb, and Lu for the 
solar photosphere and for five very metal-poor, \rpro-rich giant stars.
Combining these results with those of previous papers in this
series (cited in \S1), we have now derived very accurate 
solar/stellar abundances for the entire suite of stable RE elements.

With the single exception of Hf, the solar photospheric abundances 
agree with solar-system meteoritic values perfectly to within the 
uncertainty estimates of each.  
Our photospheric and stellar analyses have emphasized studying as many
transitions of each species as possible (up to 46 \ion{Nd}{2} lines
in the Sun, up to 72 \ion{Sm}{2} lines in \bd).
The line-to-line abundance scatters are always small when the
number of available transitions is large (typically $\sigma$~$<$~0.07).
This clearly demonstrates the reliability of the RE transition
probabilities published in this series of papers.
We argue that, with proper care in stellar analyses, trustworthy 
abundances of RE elements can be now be determined from spectra in 
which far fewer transitions are available.

Utilizing the new experimental atomic data we have determined far more 
precise stellar RE elemental abundances in five \rpro\ rich stars. 
These newly derived values show a dramatic decrease in star-to-star 
elemental abundance scatter -- all the RE elements are
now in very good (relative) agreement for these five halo stars.
Furthermore, our newly derived values indicate an almost perfect 
agreement between the average stellar abundances and the solar system 
\rpro-only abundances for a wide range of elements
in these five \rpro-rich stars.
There is no evidence for significant \spro\ contamination. 
The one exception appears to be a somewhat higher value of stellar Hf 
with respect to the solar system \rpro-only value for this element. 
This may indicate that further analysis of the solar $r$- and \spro\ 
deconvolution for this element might be useful.
These results for the five \rpro-rich halo stars confirm, and strongly 
support, early studies that indicate that the r-process was dominant 
for the \ncap\ elements early in the history of the Galaxy.

\acknowledgments %

Parts of this research were undertaken while CS was in residence
at Osservatorio Astronomico di Padova; the Director and staff
are thanked for their hospitality and financial support.
We thank Anna Frebel, Katherina Lodders, Ian Roederer, and 
Jennifer Sobeck for helpful discussions.
We appreciate the 
use of NASA's Astrophysics Data System Bibliographic Services, and 
the privilege to observe on the revered summit of Mauna Kea.  
The solar abundance analyses of the present and previous papers of this
series have greatly benefited from the availability of the
photospheric spectrum in the BASS2000 Solar Survey Archive maintained
by l'Observatoire de Paris.
This work has been supported by the National Science Foundation 
through grants AST 05-06324 to JEL and EDH, AST 06-07708 to CS, 
and AST 07-07447 to JJC.


\clearpage
\appendix

\begin{center}
APPENDIX
\end{center}

There have been numerous experimental studies of hyperfine 
structure (hfs) in \ion{Pr}{2}.   
We have reviewed the literature for measurements on the 
upper and lower levels of lines useful, or potentially useful, 
for elemental abundance studies.  
Six publications are relevant, as indicated in Table~\ref{tab10}.    
One sees generally good agreement among measured values of the hfs $A$ 
constants.   
Only a few, not very accurate, measurements of the hfs $B$ constants 
have been reported.  
Since the electric quadrupole interaction ($B$ constants) has a much 
smaller effect on the line component pattern than the magnetic dipole 
interaction ($A$ constant), it is often neglected and will be neglected here.

One of the best and fairly extensive set of measurements of \ion{Pr}{2}
hfs $A$ constants is that by Rivest \etal\ (2002)\nocite{riv02} using 
laser induced fluorescence.  
We adopted their measurements, if available, to compute the complete 
hfs line component patterns that are given in Table~\ref{tab11}.  
For levels which were not studied by Rivest \etal, we used hfs $A$ 
constants from Ginibre (1989)\nocite{gin89}.  
Iv01 improved some \ion{Pr}{2} energy levels using FTS data.  
The center-of-gravity wavenumbers in Table~\ref{tab11} are from 
the Iv01 energy levels in every case where an improved energy was 
reported for both the upper and lower level of the line.  
For other lines the center-of gravity wavenumbers are from the NIST 
energy levels (Martin \etal\ 1978)\nocite{mar78}, because it is probably 
not a good idea to mix energy levels from two sources. 
Center-of-gravity air wavelengths were computed from wavenumbers using 
the standard index of air (Edl\'en 1953).\nocite{edl53}

For \ion{Yb}{2} we used the isotopic and hyperfine data of 
M{\aa}rtensson-Pendrill, Gough, \& Hannaford (1994)\nocite{mar94}.
We adopted the transition probabilities of Bi\'emont \etal\
(1998)\nocite{bie98} renormalized to the lifetime results of Pinnington,
Rieger, \& Kernahan (1997)\nocite{pin97}: \loggf$_{3289}$~= +0.02, and
\loggf$_{3694}$~= $-$0.30.
These values are close to those derived from Bi\'emont \etal\
(2002)\nocite{bie02}, as given in the D.R.E.A.M. database\footnote{
http://w3.umh.ac.be/$\sim$astro/dream.shtml}:
\loggf$_{3289}$~= $-$0.05, and \loggf$_{3694}$~= $-$0.32.
Combining the transition probabilities, hyperfine and isotopic substructures,
and the solar isotopic breakdown given in \ref{ybtext} yields complete 
transition structures for these two \ion{Yb}{2} lines; these are listed
in Table~\ref{tab12}.

\clearpage

\clearpage

\clearpage
\nocite{iva01}\nocite{li07}
                                                                                
\clearpage

\clearpage
\nocite{del73}

\clearpage

\clearpage
\nocite{del73}

\clearpage
\nocite{and89}

\clearpage

\clearpage
                                                                                
\clearpage
                                                                                
\clearpage

\clearpage

\begin{center}

\begin{deluxetable}{lccrrrl}
\tabletypesize{\footnotesize}
\tablewidth{0pt}
\tablecaption{Spectroscopic Observations\label{tab1}}
\tablecolumns{7}
\tablehead{
\colhead{Spectrograph}                           & 
\colhead{$\lambda$ Range}                        & 
\colhead{$R$\tablenotemark{a}}                   & 
\colhead{$S/N$}                                  & 
\colhead{$F$\tablenotemark{b}}                   & 
\colhead{$\lambda_{app}$\tablenotemark{c}}       &
\colhead{Stars}                                  \\
\colhead{}                                       & 
\colhead{\AA}                                    & 
\colhead{}                                       & 
\colhead{}                                       & 
\colhead{}                                       & 
\colhead{\AA}                                    & 
\colhead{}
}
\startdata
Keck I HIRES\tablenotemark{d} 
                       & 3050--5950               & 40,000               & 100            & 1142             & 3500               & CS 31082-001, HD 221170 \\
                       &                          &                      & 150            & 1500             & 4000               & CS 31082-001, HD 221170 \\
                       &                          &                      & 200            & 1778             & 4500               & CS 31082-001, HD 221170 \\
                       & 3100--4250               & 45,000               & 100            & 1286             & 3500               & CS 22892-052 \\
                       &                          &                      & 150            & 1688             & 4000               & CS 22892-052 \\
                       &                          &                      & 200            & 2000             & 4500               & CS 22892-052 \\
                       & 3100--4650               & 45,000               & 100            & 1286             & 3500               & BD+17 3248, HD 115444 \\
                       &                          &                      & 150            & 1688             & 4000               & BD+17 3248, HD 115444 \\
                       &                          &                      & 200            & 2000             & 4500               & BD+17 3248, HD 115444 \\
McDonald ``2d-coude''\tablenotemark{e} 
                       & 3800--9000               & 60,000               & 100            & 1500             & 4000               & BD+17 3248, HD 115444 \\
                       &                          &                      & 250            & 2500             & 6000               & BD+17 3248, HD 115444 \\
                       & 3800--7800               & 60,000               &  80            & 1200             & 4000               & HD 221170 \\
                       &                          &                      & 275            & 2750             & 6000               & HD 221170 \\
                       & 4200--7000               & 60,000               &  50            &  750             & 4000               & CS 22892-052 \\
                       &                          &                      & 150            & 1500             & 6000               & CS 22892-052 \\
Magellan Clay MIKE\tablenotemark{f} 
                       & 3800--4950               & 50,000               & 100            & 1250             & 4000               & CS 22892-052 \\
                       & 5050--8000               & 38,000               & 150            &  950             & 6000               & CS 22892-052 \\
\enddata

\tablenotetext{a}{$R \equiv \lambda/\delta\lambda$}

\tablenotetext{b}{$F \equiv (R/\lambda_{app}\times(S/N))$}

\tablenotetext{c}{$\lambda_{app} \equiv$ the approximate wavelength for
calculation of $F$}

\tablenotetext{d}{Vogt \etal\ (1994); detailed description at 
http://www.ucolick.org/$\sim$hires/}

\tablenotetext{e}{Tull \etal\ (1995); detailed description at 
http://www.as.utexas.edu/mcdonald/facilities/2.7m/cs2.html}

\tablenotetext{f}{Bernstein \etal\ (2003); detailed description at 
http://www.ucolick.org/$\sim$rab/MIKE/usersguide.html}
\end{deluxetable} 

\end{center}

\clearpage

\begin{center}
\begin{deluxetable}{lccccc}
\tabletypesize{\footnotesize}
\tablewidth{0pt}
\tablecaption{Stellar Model Parameters\label{tab2}}
\tablecolumns{6}
\tablehead{
\colhead{Star}                      &
\colhead{T$_{eff}$}                 &
\colhead{log $g$}                   &
\colhead{[Fe/H]}                    &
\colhead{v$_t$}                     &
\colhead{Reference}                 \\
\colhead{}                          &
\colhead{K}                         &
\colhead{}                          &
\colhead{}                          &
\colhead{km s$^{-1}$}               &
\colhead{}                           
}
\startdata
BD+17 3248    & 5200 & 1.80 & $-$2.10 & 1.90 &  Cowan et al. (2002)  \\
CS 22892-052  & 4800 & 1.50 & $-$3.12 & 1.95 &  Sneden et al. (2003) \\
CS 31082-001  & 4825 & 1.50 & $-$2.91 & 1.90 &  Hill et al. (2002)   \\
HD 115444     & 4800 & 1.50 & $-$2.90 & 2.00 &  Westin et al. (2000) \\
HD 221170     & 4510 & 1.00 & $-$2.19 & 1.80 &  Ivans et al. (2006)  \\
\enddata

\end{deluxetable}

\end{center}

\nocite{cow02} 
\nocite{sne03}
\nocite{wes00}
\nocite{hil02}
\nocite{iva06}

\clearpage

\begin{center}
\begin{deluxetable}{lcccccccc}
\tabletypesize{\footnotesize}
\tablewidth{0pt}
\tablecaption{Solar Rare Earth Abundances\label{tab3}}
\tablecolumns{9}
\tablehead{
\colhead{Element}                                    &
\colhead{Z}                                          &
\colhead{$\log \epsilon_{met}$\tablenotemark{a}}     &
\colhead{$\log \epsilon_\odot$}                      &
\colhead{$\sigma$}                                   &
\colhead{\#\tablenotemark{b}}                        &
\colhead{Ref\tablenotemark{c}}                       &
\colhead{$\log N_r$\tablenotemark{d}}                &
\colhead{$\log N_r$\tablenotemark{d}}                \\
\colhead{}                                           &
\colhead{}                                           &
\colhead{meteoritic}                                 &
\colhead{}                                           &
\colhead{}                                           &
\colhead{}                                           &
\colhead{}                                           &
\colhead{empirical}                                  &
\colhead{stellar}                   
}
\startdata
Ba  & 56  &  2.19 $\pm$ 0.03  &     \nodata      & \nodata  & \nodata  &  1  & --0.0936  & --0.0696 \\
La  & 57  &  1.18 $\pm$ 0.06  & 1.14 $\pm$ 0.01  &    0.03  &      14  &  2  & --0.9547  & --0.9210 \\
Ce  & 58  &  1.61 $\pm$ 0.02  & 1.61 $\pm$ 0.01  &    0.06  &      45  &  3  & --0.6904  & --0.5733 \\
Pr  & 59  &  0.78 $\pm$ 0.03  & 0.76 $\pm$ 0.02  &    0.04  &       5  &  1  & --1.0862  & --1.0670 \\
Nd  & 60  &  1.46 $\pm$ 0.03  & 1.45 $\pm$ 0.01  &    0.05  &      46  &  4  & --0.3723  & --0.5163 \\
Sm  & 62  &  0.95 $\pm$ 0.04  & 1.00 $\pm$ 0.01  &    0.05  &      36  &  5  & --0.7595  & --0.7592 \\
Eu  & 63  &  0.52 $\pm$ 0.04  & 0.52 $\pm$ 0.01  &    0.04  &      14  &  6  & --1.0424  & --1.0376 \\
Gd  & 64  &  1.06 $\pm$ 0.02  & 1.11 $\pm$ 0.01  &    0.05  &      20  &  7  & --0.5591  & --0.5546 \\
Tb  & 65  &  0.31 $\pm$ 0.03  & 0.28 $\pm$ 0.07  &    0.10  &       2  &  8  & --1.2218  & --1.2526 \\
Dy  & 66  &  1.13 $\pm$ 0.04  & 1.13 $\pm$ 0.02  &    0.06  &      13  &  1  & --0.4437  & --0.4755 \\
Ho  & 67  &  0.49 $\pm$ 0.02  & 0.51 $\pm$ 0.10  &    0.10  &       3  &  9  & --1.0899  & --1.0862 \\
Er  & 68  &  0.95 $\pm$ 0.03  & 0.96 $\pm$ 0.03  &    0.06  &       8  & 10  & --0.6798  & --0.6832 \\
Tm  & 69  &  0.11 $\pm$ 0.06  & 0.14 $\pm$ 0.02  &    0.04  &       3  &  1  & --1.5086  & --1.4841 \\
Yb  & 70  &  0.94 $\pm$ 0.03  & 0.86 $\pm$ 0.10  &    0.10  &       1  &  1  & --0.7889  & --0.7783 \\
Lu  & 71  &  0.09 $\pm$ 0.06  & 0.12 $\pm$ 0.08  &    0.08  &       1  &  1  & --1.5100  & --1.5317 \\
Hf  & 72  &  0.77 $\pm$ 0.04  & 0.88 $\pm$ 0.02  &    0.03  &       4  & 11  & --1.0974  & --1.1675 \\
\enddata

\tablenotetext{a}{Lodders 2003}

\tablenotetext{b}{number of lines used for the photospheric abundance}

\tablenotetext{c}{reference for the photospheric abundance} 

\tablenotetext{d}{Estimates of the $r$-process only abundances $N_r$ of 
                  solar-system RE elements, based on the differences
                  between total meteoritic abundances $N_{met}$ and
                  ``empirical'' and ``stellar'' estimates of the 
                  $s$-process only abundances $N_s$; see text for explanation
                  of these estimates.  These meteoritic abundances 
                  (normalized to $\log N({\rm Si})$ = 6) can be 
                  translated to photospheric ones (normalized to 
                  $\log \epsilon({\rm H})$ = 12) through
                  $\log \epsilon$ = $\log N$ + 1.54}
\tablerefs{
1.  This paper;  
2.  Lawler \etal\ (2001a);
3.  Lawler \etal\ (2009);
4.  Den Hartog \etal\ (2003);
5.  Lawler \etal\ (2006);
6.  Lawler \etal\ (2001c);
7.  Den Hartog \etal\ (2006);
8.  Lawler \etal\ (2001b);
9.  Lawler \etal\ (2004);
10. Lawler \etal\ (2008b);
11. Lawler \etal\ (2007)
}

\end{deluxetable}

\end{center}

\nocite{law01a} 
\nocite{law09}
\nocite{den03} 
\nocite{law06}
\nocite{law01c} 
\nocite{den06} 
\nocite{law01b} 
\nocite{law04}
\nocite{law08b} 
\nocite{law07}

\nocite{lod03}

\clearpage

\begin{center}
\begin{deluxetable}{lccccccccccccc}
\tabletypesize{\scriptsize}
\tablewidth{0pt}
\tablecaption{Rare Earth Abundances for BD+17 3248, CS 22892-052, 
              and CS 31082-001\label{tab4}}
\tablecolumns{14}
\tablehead{
\colhead{}                          &
\colhead{}                          &
\multicolumn{4}{c}{BD+17 3248}      & 
\multicolumn{4}{c}{CS 22892-052}    & 
\multicolumn{4}{c}{CS 31082-001}    \\
\colhead{El}                        &
\colhead{Z}                         &
\colhead{$\log \epsilon$}           &
\colhead{$\sigma$}                  &
\colhead{\#\tablenotemark{a}}       &
\colhead{Ref\tablenotemark{b}}      &
\colhead{$\log \epsilon$}           &
\colhead{$\sigma$}                  &
\colhead{\#\tablenotemark{a}}       &
\colhead{Ref\tablenotemark{b}}      &
\colhead{$\log \epsilon$}           &
\colhead{$\sigma$}                  &
\colhead{\#\tablenotemark{a}}       &
\colhead{Ref\tablenotemark{b}}      
}
\startdata
Ba & 56  &
       +0.48 $\pm$  0.05  &  0.11  &     4  &     1  &
      --0.01 $\pm$  0.06  &  0.12  &     4  &     1  &
     \nodata              & \nodata& \nodata&     1  \\
La & 57  &
      --0.42 $\pm$  0.01  &  0.05  &    15  &     2  &
      --0.84 $\pm$  0.01  &  0.05  &    15  &     2  &
      --0.62 $\pm$  0.01  &  0.04  &     9  &     2  \\
Ce & 58  &
      --0.11 $\pm$  0.01  &  0.05  &    40  &     4  &
      --0.46 $\pm$  0.01  &  0.05  &    32  &     4  &
      --0.29 $\pm$  0.01  &  0.03  &    38  &     4  \\
Pr & 59  &
      --0.71 $\pm$  0.02  &  0.06  &    18  &     1  &
      --0.96 $\pm$  0.02  &  0.07  &    15  &     1  &
      --0.79 $\pm$  0.01  &  0.07  &    27  &     1  \\
Nd & 60  &
      --0.09 $\pm$  0.01  &  0.06  &    57  &     5  &
      --0.37 $\pm$  0.01  &  0.06  &    37  &     5  &
      --0.15 $\pm$  0.01  &  0.08  &    68  &     1  \\
Sm & 62  &
      --0.34 $\pm$  0.01  &  0.05  &    72  &     6  &
      --0.61 $\pm$  0.01  &  0.07  &    55  &     6  &
      --0.42 $\pm$  0.01  &  0.04  &    67  &     1  \\
Eu & 63  &
      --0.68 $\pm$  0.01  &  0.04  &     9  &     1  &
      --0.95 $\pm$  0.01  &  0.02  &     9  &     1  &
      --0.72 $\pm$  0.01  &  0.03  &     7  &     1  \\
Gd & 64  &
      --0.14 $\pm$  0.01  &  0.04  &    41  &     7  &
      --0.42 $\pm$  0.01  &  0.07  &    32  &     7  &
      --0.21 $\pm$  0.01  &  0.05  &    32  &     1  \\
Tb & 65  &
      --0.91 $\pm$  0.02  &  0.05  &     5  &     8  &
      --1.13 $\pm$  0.01  &  0.04  &     7  &     9  &
      --1.01 $\pm$  0.01  &  0.04  &     9  &     1  \\
Dy & 66  &
      --0.04 $\pm$  0.01  &  0.05  &    28  &     1  &
      --0.26 $\pm$  0.01  &  0.06  &    29  &     1  &
      --0.07 $\pm$  0.01  &  0.05  &    35  &     1  \\
Ho & 67  &
      --0.70 $\pm$  0.02  &  0.05  &    11  &    10  &
      --0.92 $\pm$  0.01  &  0.02  &    13  &    10  &
      --0.80 $\pm$  0.03  &  0.09  &    12  &     1  \\
Er & 68  &
      --0.25 $\pm$  0.01  &  0.04  &    17  &    11  &
      --0.48 $\pm$  0.01  &  0.04  &    21  &    11  &
      --0.30 $\pm$  0.01  &  0.04  &    19  &    11  \\
Tm & 69  &
      --1.12 $\pm$  0.02  &  0.05  &     6  &     1  &
      --1.39 $\pm$  0.02  &  0.04  &     6  &     1  &
      --1.15 $\pm$  0.02  &  0.06  &     7  &     1  \\
Yb & 70  &
      --0.27 $\pm$  0.10  &  0.10  &     1  &     1  &
      --0.55 $\pm$  0.10  &  0.10  &     1  &     1  &
      --0.41 $\pm$  0.10  &  0.10  &     1  &     1  \\
Lu & 71  &
     \nodata              & \nodata& \nodata&     1  &
     \nodata              & \nodata& \nodata&     1  &
     \nodata              & \nodata& \nodata&     1  \\
Hf & 72  &
      --0.57 $\pm$  0.03  &  0.08  &     6  &     2  &
      --0.88 $\pm$  0.01  &  0.04  &     8  &     2  &
      --0.72 $\pm$  0.01  &  0.04  &    10  &     2  \\
\enddata

\tablenotetext{a}{number of lines used for the stellar abundance}

\tablenotetext{b}{reference for the stellar abundance; these are
                  cited at the end of Table~\ref{tab5}}

\end{deluxetable}

\end{center}

\clearpage

\begin{center}
\begin{deluxetable}{lccccccccc}
\tabletypesize{\scriptsize}
\tablewidth{0pt}
\tablecaption{Rare Earth Abundances for HD 115444 and 
              HD 221170\label{tab5}}
\tablecolumns{10}
\tablehead{
\colhead{}                          &
\colhead{}                          &
\multicolumn{4}{c}{HD 115444}       & 
\multicolumn{4}{c}{HD 221170}       \\
\colhead{El}                        &
\colhead{Z}                         &
\colhead{$\log \epsilon$}           &
\colhead{$\sigma$}                  &
\colhead{\#\tablenotemark{a}}       &
\colhead{Ref\tablenotemark{b}}      &
\colhead{$\log \epsilon$}           &
\colhead{$\sigma$}                  &
\colhead{\#\tablenotemark{a}}       &
\colhead{Ref\tablenotemark{b}}      
}
\startdata
Ba & 56  &
      --0.73 $\pm$  0.04  &  0.08  &     4  &     1  &
       +0.18 $\pm$  0.05  &  0.11  &     4  &     1    \\
La & 57  &
      --1.44 $\pm$  0.02  &  0.05  &     8  &     2  &
      --0.73 $\pm$  0.01  &  0.06  &    36  &     3    \\
Ce & 58  &
      --1.06 $\pm$  0.01  &  0.07  &    26  &     4  &
      --0.42 $\pm$  0.01  &  0.04  &    37  &     4    \\
Pr & 59  &
      --1.57 $\pm$  0.02  &  0.06  &    10  &     1  &
      --1.00 $\pm$  0.02  &  0.07  &    21  &     1    \\
Nd & 60  &
      --1.02 $\pm$  0.01  &  0.08  &    37  &     5  &
      --0.35 $\pm$  0.01  &  0.08  &    63  &     3    \\
Sm & 62  &
      --1.26 $\pm$  0.01  &  0.07  &    67  &     6  &
      --0.66 $\pm$  0.01  &  0.07  &    28  &     3    \\
Eu & 63  &
      --1.64 $\pm$  0.02  &  0.04  &     8  &     1  &
      --0.89 $\pm$  0.03  &  0.07  &     7  &     1    \\
Gd & 64  &
      --1.08 $\pm$  0.01  &  0.07  &    29  &     7  &
      --0.46 $\pm$  0.04  &  0.14  &    11  &     3    \\
Tb & 65  &
      --1.84 $\pm$  0.04  &  0.08  &     3  &     1  &
      --1.21 $\pm$  0.03  &  0.08  &     8  &     3    \\
Dy & 66  &
      --1.00 $\pm$  0.01  &  0.07  &    24  &     1  &
      --0.29 $\pm$  0.01  &  0.06  &    25  &     1    \\
Ho & 67  &
      --1.61 $\pm$  0.01  &  0.04  &     9  &    10  &
      --0.97 $\pm$  0.02  &  0.07  &     8  &     3    \\
Er & 68  &
      --1.22 $\pm$  0.02  &  0.07  &    15  &    11  &
      --0.47 $\pm$  0.02  &  0.08  &    14  &    11    \\
Tm & 69  &
      --2.06 $\pm$  0.02  &  0.04  &     5  &     1  &
      --1.39 $\pm$  0.03  &  0.06  &     6  &     1    \\
Yb & 70  &
      --1.43 $\pm$  0.10  &  0.10  &     1  &     1  &
      --0.48 $\pm$  0.10  &  0.10  &     1  &     1    \\
Lu & 71  &
     \nodata              & \nodata& \nodata&     1  &
     \nodata              & \nodata& \nodata&     1    \\
Hf & 72  &
      --1.51 $\pm$  0.01  &  0.03  &     4  &     2  &
      --0.84 $\pm$  0.03  &  0.11  &    10  &     2    \\
\enddata

\tablenotetext{a}{number of lines used for the stellar abundance}

\tablenotetext{b}{reference for the stellar abundance}

\tablerefs{
1.  This paper;  
2.  Lawler \etal\ (2007);
3.  Ivans \etal\ (2006);
4.  Lawler \etal\ (2009);
5.  Den Hartog \etal\ (2003);
6.  Lawler \etal\ (2006);
7.  Den Hartog \etal\ (2006);
8.  Cowan \etal\ (2002);
9.  Sneden \etal\ (2003);
10. Lawler \etal\ (2004);
11. Lawler \etal\ (2008b);
}

\end{deluxetable}

\end{center}

\nocite{law07}
\nocite{iva06}
\nocite{law09}
\nocite{den03} 
\nocite{law06}
\nocite{den06} 
\nocite{cow02}
\nocite{sne03}
\nocite{law04}
\nocite{law08b}

\clearpage

\begin{center}
\begin{deluxetable}{lccccccccc}
\tabletypesize{\footnotesize}
\tablewidth{0pt}
\tablecaption{\ion{Pr}{2} Line Abundances\label{tab6}}
\tablecolumns{10}
\tablehead{
\colhead{$\lambda$}           &
\colhead{$\chi$}              &
\colhead{$\log gf$}           &
\colhead{$\log gf$}           &
\colhead{$\log \epsilon$}     &
\colhead{$\log \epsilon$}     &
\colhead{$\log \epsilon$}     &
\colhead{$\log \epsilon$}     &
\colhead{$\log \epsilon$}     &
\colhead{$\log \epsilon$}     \\
\colhead{\AA}                 &
\colhead{eV}                  &
\colhead{Li07}                &
\colhead{Iv01}                &
\colhead{$\odot$}             &
\colhead{BD+17 3248}          &
\colhead{CS 22892-052}        &
\colhead{CS 31082-001}        &
\colhead{HD 115444}           &
\colhead{HD 221170}           
}
\startdata
3964.82  & 0.055  &  +0.069  &  +0.121  & \nodata  &  --0.75  &  --1.00  & \nodata  & \nodata  &  --1.03 \\
3965.26  & 0.204  &  +0.204  &  +0.135  & \nodata  &  --0.85  &  --1.03  & \nodata  & \nodata  &  --1.03 \\
4004.70  & 0.216  & --0.250  & \nodata  & \nodata  & \nodata  &  --0.90  &  --0.71  & \nodata  & \nodata \\
4015.39  & 0.216  & --0.362  & \nodata  & \nodata  & \nodata  & \nodata  &  --0.84  & \nodata  & \nodata \\
4039.34  & 0.204  & --0.336  & \nodata  & \nodata  & \nodata  & \nodata  &  --0.76  & \nodata  &  --0.98 \\
4044.81  & 0.000  & --0.293  & \nodata  & \nodata  &  --0.65  &  --0.92  &  --0.72  & \nodata  &  --0.88 \\
4062.81  & 0.422  &  +0.334  & \nodata  & \nodata  &  --0.59  &  --0.83  &  --0.63  & \nodata  &  --0.83 \\
4096.82  & 0.216  & --0.255  & \nodata  & \nodata  &  --0.75  & \nodata  &  --0.81  & \nodata  & \nodata \\
4118.46  & 0.055  &  +0.175  & \nodata  & \nodata  & \nodata  &  --0.85  &  --0.68  & \nodata  & \nodata \\
4141.22  & 0.550  &  +0.381  & \nodata  & \nodata  &  --0.80  &  --1.04  &  --0.86  & \nodata  &  --1.08 \\
4143.12  & 0.371  &  +0.604  &  +0.609  & \nodata  &  --0.68  & \nodata  &  --0.71  &  --1.49  & \nodata \\
4164.16  & 0.204  &  +0.170  &  +0.160  & \nodata  &  --0.75  &  --1.00  &  --0.84  & \nodata  &  --1.05 \\
4179.40  & 0.204  &  +0.459  &  +0.477  & \nodata  &  --0.58  &  --0.98  &  --0.79  &  --1.49  &  --0.88 \\
4189.48  & 0.371  &  +0.431  &  +0.382  & \nodata  &  --0.72  &  --1.02  &  --0.86  &  --1.64  &  --1.03 \\
4222.95  & 0.055  &  +0.235  &  +0.271  &   +0.71  &  --0.70  &  --1.00  &  --0.74  &  --1.61  &  --1.00 \\
4405.83  & 0.550  & --0.062  & --0.037  & \nodata  & \nodata  & \nodata  &  --0.71  & \nodata  & \nodata \\
4408.82  & 0.000  &  +0.053  &  +0.179  & \nodata  &  --0.70  & \nodata  &  --0.71  &  --1.53  &  --0.94 \\
4413.77  & 0.216  & --0.563  & \nodata  & \nodata  & \nodata  & \nodata  &  --0.73  & \nodata  & \nodata \\
4429.13  & 0.000  & --0.495  & \nodata  & \nodata  &  --0.70  &  --1.02  &  --0.78  &  --1.59  &  --1.03 \\
4429.26  & 0.371  & --0.048  & --0.103  & \nodata  &          &          &          &          &         \\
4449.83  & 0.204  & --0.261  & --0.174  & \nodata  &  --0.70  & \nodata  &  --0.76  & \nodata  &  --0.97 \\
4496.33  & 0.055  & --0.368  & --0.268  & \nodata  &  --0.72  &  --0.90  &  --0.76  &  --1.49  &  --0.97 \\
4496.47  & 0.216  & --0.762  & \nodata  & \nodata  &          &          &          &          &         \\
4510.15  & 0.422  & --0.007  & --0.023  &   +0.78  &  --0.72  & \nodata  &  --0.86  &  --1.64  &  --1.02 \\
5129.54  & 0.648  & --0.134  & \nodata  & \nodata  & \nodata  & \nodata  &  --0.81  & \nodata  &  --1.01 \\
5135.15  & 0.949  & \nodata  &  +0.008  & \nodata  & \nodata  & \nodata  &  --0.91  & \nodata  &  --1.03 \\
5173.91  & 0.967  &  +0.359  &  +0.384  & \nodata  & \nodata  & \nodata  &  --0.86  & \nodata  & \nodata \\
5219.05  & 0.795  & (+0.405)\tablenotemark{a}
                             & --0.053  &   +0.81  & \nodata  & \nodata  & \nodata  & \nodata  & \nodata \\
5220.11  & 0.795  & \nodata  &  +0.298  & \nodata  &  --0.72  &  --1.00  &  --0.89  &  --1.59  &  --1.08 \\
5259.73  & 0.633  & \nodata  &  +0.114  &   +0.78  &  --0.70  &  --0.91  &  --0.86  &  --1.64  &  --1.08 \\
5292.62  & 0.648  & --0.269  & --0.257  & \nodata  & \nodata  & \nodata  &  --0.83  & \nodata  &  --1.05 \\
5322.77  & 0.482  & --0.123  & --0.319  &   +0.74  & \nodata  & \nodata  &  --0.81  & \nodata  &  --1.05 \\
5352.40  & 0.482  & --0.739  & \nodata  & \nodata  & \nodata  & \nodata  & \nodata  & \nodata  & \nodata \\
\enddata

\tablenotetext{a}{Li07 note that this is a blended line in their spectrum;
                  we used the $\log gf$ from Iv01.}

\end{deluxetable}

\end{center}

\clearpage

\begin{center}
\begin{deluxetable}{lcccccccc}
\tabletypesize{\footnotesize}
\tablewidth{0pt}
\tablecaption{\ion{Dy}{2} Line Abundances\label{tab7}}
\tablecolumns{9}
\tablehead{
\colhead{$\lambda$}           &
\colhead{$\chi$}              &
\colhead{$\log gf$}           &
\colhead{$\log \epsilon$}     &
\colhead{$\log \epsilon$}     &
\colhead{$\log \epsilon$}     &
\colhead{$\log \epsilon$}     &
\colhead{$\log \epsilon$}     &
\colhead{$\log \epsilon$}     \\
\colhead{\AA}                 &
\colhead{eV}                  &
\colhead{}                    &
\colhead{$\odot$}             &
\colhead{BD+17 3248}          &
\colhead{CS 22892-052}        &
\colhead{CS 31082-001}        &
\colhead{HD 115444}           &
\colhead{HD 221170}           
}
\startdata
3407.80  & 0.000  &  +0.18  & \nodata  &  --0.04  &  --0.20  &  --0.02  &  --1.04  & \nodata \\
3413.78  & 0.103  & --0.52  & \nodata  &  --0.01  &  --0.23  &  --0.07  &  --1.01  & \nodata \\
3434.37  & 0.000  & --0.45  &   +1.00  &  --0.09  &  --0.23  &  --0.05  &  --0.90  & \nodata \\
3454.32  & 0.103  & --0.14  &   +1.20  &  --0.09  &  --0.33  &  --0.17  & \nodata  &  --0.28 \\
3456.56  & 0.589  & --0.11  & \nodata  &  --0.01  &  --0.35  &  --0.15  &  --0.90  &  --0.28 \\
3460.97  & 0.000  & --0.07  & \nodata  &  --0.19  &  --0.28  &  --0.13  &  --1.04  &  --0.28 \\
3523.98  & 0.538  &  +0.42  & \nodata  &  --0.10  & \nodata  &  --0.12  &  --1.04  & \nodata \\
3531.71  & 0.000  &  +0.77  &   +1.20  &  --0.02  &  --0.23  &  --0.03  &  --1.11  &  --0.18 \\
3534.96  & 0.103  & --0.04  & \nodata  &  --0.11  &  --0.35  &  --0.10  & \nodata  & \nodata \\
3536.02  & 0.538  &  +0.53  &   +1.10  &  --0.07  &  --0.35  &  --0.05  &  --1.11  &  --0.23 \\
3546.83  & 0.103  & --0.55  &   +1.23  &  --0.01  &  --0.23  &  --0.07  &  --0.91  &  --0.30 \\
3550.22  & 0.589  &  +0.27  & \nodata  &  --0.04  &  --0.35  &  --0.17  &  --1.06  &  --0.31 \\
3551.62  & 0.589  &  +0.02  & \nodata  & \nodata  &  --0.32  &  --0.15  & \nodata  & \nodata \\
3563.15  & 0.103  & --0.36  & \nodata  &  --0.06  &  --0.30  &  --0.13  &  --1.05  &  --0.30 \\
3630.24  & 0.538  &  +0.04  & \nodata  &  --0.05  &  --0.30  &  --0.09  &  --0.94  &  --0.27 \\
3630.48  & 0.925  & --0.66  & \nodata  & \nodata  & \nodata  &  --0.07  & \nodata  & \nodata \\
3694.81  & 0.103  & --0.11  &   +1.11  &  --0.06  &  --0.28  &  --0.08  &  --1.03  &  --0.25 \\
3747.82  & 0.103  & --0.81  & \nodata  &  --0.05  &  --0.28  &  --0.08  &  --0.92  &  --0.34 \\
3757.37  & 0.103  & --0.17  & \nodata  &  --0.05  &  --0.31  &  --0.08  &  --1.01  &  --0.28 \\
3788.44  & 0.103  & --0.57  & \nodata  &  --0.02  &  --0.23  &  --0.07  &  --0.96  &  --0.35 \\
3944.68  & 0.000  &  +0.11  & \nodata  &  --0.06  &  --0.20  &   +0.00  &  --1.06  &  --0.30 \\
3978.56  & 0.925  &  +0.22  & \nodata  & \nodata  &  --0.30  & \nodata  & \nodata  & \nodata \\
3983.65  & 0.538  & --0.31  &   +1.08  &  --0.02  &  --0.28  &  --0.07  &  --0.97  &  --0.26 \\
3996.69  & 0.589  & --0.26  &   +1.10  &  --0.03  &  --0.26  &  --0.08  &  --0.96  &  --0.36 \\
4011.29  & 0.925  & --0.73  &   +1.14  & \nodata  & \nodata  &  --0.08  & \nodata  & \nodata \\
4014.70  & 0.927  & --0.70  & \nodata  & \nodata  &  --0.23  &  --0.03  & \nodata  & \nodata \\
4041.98  & 0.927  & --0.90  & \nodata  & \nodata  & \nodata  &  --0.05  & \nodata  & \nodata \\
4050.57  & 0.589  & --0.47  & \nodata  &  --0.03  &  --0.23  &  --0.05  &  --1.01  &  --0.38 \\
4073.12  & 0.538  & --0.32  &   +1.10  &  --0.02  &  --0.23  &  --0.07  &  --1.06  &  --0.41 \\
4077.97  & 0.103  & --0.04  &   +1.17  &  --0.01  &  --0.16  &  --0.03  &  --0.94  &  --0.28 \\
4103.31  & 0.103  & --0.38  &   +1.17  &   +0.11  &  --0.15  &   +0.02  &  --0.91  &  --0.13 \\
4124.63  & 0.925  & --0.66  & \nodata  &  --0.04  &  --0.26  &  --0.02  & \nodata  &  --0.30 \\
4409.38  & 0.000  & --1.24  & \nodata  &   +0.03  & \nodata  &  --0.02  & \nodata  &  --0.28 \\
4449.70  & 0.000  & --1.03  &   +1.12  &   +0.06  &  --0.22  &  --0.05  &  --0.96  &  --0.38 \\
4620.04  & 0.103  & --1.93  & \nodata  & \nodata  & \nodata  &   +0.05  & \nodata  &  --0.28 \\
5169.69  & 0.103  & --1.95  & \nodata  & \nodata  & \nodata  &  --0.05  & \nodata  &  --0.33 \\
\enddata

\end{deluxetable}

\end{center}

\clearpage

\begin{center}
\begin{deluxetable}{lcccccccc}
\tabletypesize{\footnotesize}
\tablewidth{0pt}
\tablecaption{\ion{Tm}{2} Line Abundances\label{tab8}}
\tablecolumns{9}
\tablehead{
\colhead{$\lambda$}           &
\colhead{$\chi$}              &
\colhead{$\log gf$}           &
\colhead{$\log \epsilon$}     &
\colhead{$\log \epsilon$}     &
\colhead{$\log \epsilon$}     &
\colhead{$\log \epsilon$}     &
\colhead{$\log \epsilon$}     &
\colhead{$\log \epsilon$}     \\
\colhead{\AA}                 &
\colhead{eV}                  &
\colhead{}                    &
\colhead{$\odot$}             &
\colhead{BD+17 3248}          &
\colhead{CS 22892-052}        &
\colhead{CS 31082-001}        &
\colhead{HD 115444}           &
\colhead{HD 221170}           
}
\startdata
3240.23  & 0.029  & --0.80  & \nodata  & \nodata  & \nodata  &  --1.09  & \nodata  & \nodata \\
3462.20  & 0.000  &  +0.03  &   +0.18  &  --1.20  &  --1.37  &  --1.24  &  --2.13  &  --1.37 \\
3700.26  & 0.029  & --0.38  &   +0.13  &  --1.06  &  --1.34  &  --1.11  &  --2.04  &  --1.35 \\
3701.36  & 0.000  & --0.54  &   +0.10  &  --1.10  &  --1.42  &  --1.17  &  --2.04  &  --1.35 \\
3795.76  & 0.029  & --0.23  & \nodata  &  --1.13  &  --1.44  &  --1.19  &  --2.07  &  --1.52 \\
3848.02  & 0.000  & --0.14  & \nodata  &  --1.12  &  --1.37  &  --1.19  &  --2.02  &  --1.37 \\
3996.51  & 0.000  & --1.20  & \nodata  &  --1.10  &  --1.37  &  --1.07  & \nodata  &  --1.40 \\
\enddata

\end{deluxetable}

\end{center}

\clearpage

\begin{center}
\begin{deluxetable}{crrrrrr}
\tabletypesize{\footnotesize}
\tablewidth{0pt}
\tablecaption{Abundance Sensitivities to Parameter Changes\label{tab9}}
\tablecolumns{7}
\tablehead{
\colhead{parameter=}                &
\colhead{T$_{eff}$}                 &
\colhead{$\log g$}                  &
\colhead{$v_t$}                     &
\colhead{[M/H]}                     &
\colhead{scat\tablenotemark{a}}     &
\colhead{model}                     \\
\colhead{change\tablenotemark{b}}   &
\colhead{$+$150}                    &
\colhead{$+$0.5}                    &
\colhead{$+$0.5}                    &
\colhead{$+$0.5}                    &
\colhead{yes}                       &
\colhead{MARCS\tablenotemark{c}}
}
\startdata
La       & $+$0.10 & $+$0.16 & $-$0.02 & $+$0.03 & $-$0.07 & $-$0.01 \\
Ce       & $+$0.09 & $+$0.14 & $-$0.02 & $+$0.04 & $-$0.05 & $-$0.01 \\
Pr       & $+$0.11 & $+$0.14 & $-$0.02 & $+$0.04 & $-$0.05 &    0.00 \\
Eu       & $+$0.11 & $+$0.16 & $-$0.03 & $+$0.03 & $-$0.05 & $+$0.01 \\
Dy       & $+$0.10 & $+$0.12 & $-$0.05 & $+$0.03 & $-$0.10 & $-$0.01 \\
Er       & $+$0.10 & $+$0.11 & $-$0.08 & $+$0.03 & $-$0.12 & $-$0.01 \\
Tm       & $+$0.09 & $+$0.13 & $-$0.03 & $+$0.04 & $-$0.10 &    0.00 \\
Yb       & $+$0.08 & $+$0.05 & $-$0.20 & $-$0.05 & $-$0.23 & $-$0.06 \\
\\
                                                                
$<>$     & $+$0.10 & $+$0.13 & $-$0.06 & $+$0.02 & $-$0.10 & $-$0.01 \\
$\sigma$ &    0.01 &    0.04 &    0.06 &    0.03 &    0.06 &    0.03 \\
\enddata

\tablenotetext{a}{Continuum source function computed with scattering.}

\tablenotetext{b}{Change of a model parameter from a baseline model taken
from Kurucz (1998) grid computed with parameters T$_{eff}$~=~4750~K,
$\log g$~=~1.5, $v_t$~=~2.0~km~s$^{-1}$, [M/H]~=~$-$2.5, and no
correction for scattering opacity in the continuum source function.}

\tablenotetext{c}{Gustafsson \etal\ (2008).\nocite{gus08}}

\end{deluxetable}

\end{center}

\clearpage
 
\begin{center}
\begin{deluxetable}{ccccc}
\tabletypesize{\footnotesize}
\tablewidth{0pt}
\tablecaption{Measured hfs $A$ Constants for \ion{Pr}{2} Levels of Interest 
in This Investigation\label{tab10}}
\tablecolumns{5}
\tablehead{
\colhead{Energy$^a$}          &
\colhead{Energy$^c$}          &
\colhead{$J$}                 &
\colhead{hfs $A$}             &
\colhead{Ref.}                \\
\colhead{cm$^{-1}$}           &
\colhead{cm$^{-1}$}           &
\colhead{}                    &
\colhead{0.001 cm$^{-1}$}     &
\colhead{} 
}
\startdata
    0.00 &     0.000 & 4 & --7.962 $\pm$ 0.013 & b \\
         &           &   & --7.3   $\pm$ 0.9   & c \\
         &           &   & --7.3               & d \\
  441.95 &   442.079 & 5 &  63.721 $\pm$ 0.070 & b \\
         &           &   &  63.8   $\pm$ 1.1   & c \\
         &           &   &  63.9               & d \\
 1649.01 &  1649.092 & 6 &  54.498 $\pm$ 0.090 & b \\
         &           &   &  53.8   $\pm$ 1.8   & c \\
         &           &   &  54.5               & d \\
 1743.72 &  1743.776 & 5 &  -1.478 $\pm$ 0.030 & b \\
         &           &   &  -1.6   $\pm$ 0.5   & c \\
         &           &   &  -1.3               & d \\
\enddata

\tablerefs{
a. Martin \etal\ (1978);
b. Rivest \etal\ (2002);
c. Iva01;
d. Ginibre (1989);
e. Li \etal\ (2000b);
f. Li \etal\ (2000a);
g. Ma \etal\ (1999)
}
                                                                                
\tablecomments{This table is presented in its entirety as a formatted
ASCII file in the electronic edition of the Astrophysical Journal Supplement.
A portion is shown here for guidance regarding its form and content.}
                                                                                
\end{deluxetable}
                                                                                
\end{center}
                                                                                
\nocite{mar78}
\nocite{riv02}
\nocite{gin89}
\nocite{li00b}
\nocite{li00a}
\nocite{ma99}

\clearpage
                                                                                
\begin{center}
\begin{deluxetable}{ccccccc}
\tabletypesize{\footnotesize}
\tablewidth{0pt}
\tablecaption{Hyperfine Structure Line Component Patterns for 
\ion{Pr}{2}\label{tab11}}
\tablecolumns{7}
\tablehead{
\colhead{Wavenumber\tablenotemark{a}}      &
\colhead{$\lambda_{air}$\tablenotemark{a}} &
\colhead{$F_{upp}$}                        &
\colhead{$F_{low}$}                        &
\colhead{Component}                        &
\colhead{Component}                        &
\colhead{Strength\tablenotemark{c}}        \\
\colhead{}                                 &
\colhead{}                                 &
\colhead{}                                 &
\colhead{}                                 &
\colhead{Position\tablenotemark{b}}        &
\colhead{Position\tablenotemark{b}}        &
\colhead{}                                 \\
\colhead{cm$^{-1}$}                        &
\colhead{\AA}                              &
\colhead{}                                 &
\colhead{}                                 &
\colhead{cm$^{-1}$}                        &
\colhead{\AA}                              &
\colhead{}
}
\startdata
25467.565 & 3925.4515 & 6.5 & 6.5 &   0.32402 & --0.049944 & 0.23932 \\
25467.565 & 3925.4515 & 6.5 & 5.5 &   0.27227 & --0.041967 & 0.01994 \\
25467.565 & 3925.4515 & 5.5 & 6.5 &   0.16516 & --0.025458 & 0.01994 \\
25467.565 & 3925.4515 & 5.5 & 5.5 &   0.11341 & --0.017481 & 0.17164 \\
25467.565 & 3925.4515 & 5.5 & 4.5 &   0.06962 & --0.010731 & 0.03064 \\
25467.565 & 3925.4515 & 4.5 & 5.5 & --0.02101 &   0.003239 & 0.03064 \\
25467.565 & 3925.4515 & 4.5 & 4.5 & --0.06480 &   0.009989 & 0.12121 \\
25467.565 & 3925.4515 & 4.5 & 3.5 & --0.10063 &   0.015512 & 0.03333 \\
25467.565 & 3925.4515 & 3.5 & 4.5 & --0.17479 &   0.026941 & 0.03333 \\
25467.565 & 3925.4515 & 3.5 & 3.5 & --0.21062 &   0.032464 & 0.08571 \\
25467.565 & 3925.4515 & 3.5 & 2.5 & --0.23848 &   0.036760 & 0.02910 \\
25467.565 & 3925.4515 & 2.5 & 3.5 & --0.29616 &   0.045650 & 0.02910 \\
25467.565 & 3925.4515 & 2.5 & 2.5 & --0.32402 &   0.049945 & 0.06349 \\
25467.565 & 3925.4515 & 2.5 & 1.5 & --0.34393 &   0.053014 & 0.01852 \\
25467.565 & 3925.4515 & 1.5 & 2.5 & --0.38512 &   0.059364 & 0.01852 \\
25467.565 & 3925.4515 & 1.5 & 1.5 & --0.40503 &   0.062432 & 0.05556 \\
\enddata

\tablenotetext{a}{Center-of-gravity value}

\tablenotetext{b}{Relative to the center-of-gravity value}

\tablenotetext{c}{Normalized to 1 for the whole transition}

\tablecomments{This table is presented in its entirety as a formatted
ASCII file in the electronic edition of the Astrophysical Journal Supplement.
A portion is shown here for guidance regarding its form and content.}

\end{deluxetable}

\end{center}

\clearpage
                                                                                
\begin{center}
\begin{deluxetable}{ccccrrrc}
\tabletypesize{\footnotesize}
\tablewidth{0pt}
\tablecaption{Isotopic and Hyperfine Structure Line Component Patterns for 
\ion{Yb}{2}\label{tab12}}
\tablecolumns{8}
\tablehead{
\colhead{Wavenumber\tablenotemark{a}}      &
\colhead{$\lambda_{air}$\tablenotemark{a}} &
\colhead{$F_{upp}$}                        &
\colhead{$F_{low}$}                        &
\colhead{Component}                        &
\colhead{Component}                        &
\colhead{Strength\tablenotemark{c}}        \\
\colhead{}                                 &
\colhead{}                                 &
\colhead{}                                 &
\colhead{}                                 &
\colhead{Position\tablenotemark{b}}        &
\colhead{Position\tablenotemark{b}}        &
\colhead{}                                 &
\colhead{Isotope}                          \\
\colhead{cm$^{-1}$}                        &
\colhead{\AA}                              &
\colhead{}                                 &
\colhead{}                                 &
\colhead{cm$^{-1}$}                        &
\colhead{\AA}                              &
\colhead{}                                 &
\colhead{}
}
\startdata
30392.23 & 3289.367 & 1.5 & 0.5 &   0.09884 & --0.010697 & 0.00130 & 168 \\
30392.23 & 3289.367 & 1.5 & 0.5 &   0.06745 & --0.007300 & 0.03040 & 170 \\
30392.23 & 3289.367 & 1.5 & 0.5 &   0.01878 & --0.002033 & 0.21830 & 172 \\
30392.23 & 3289.367 & 1.5 & 0.5 & --0.01971 &   0.002134 & 0.31830 & 174 \\
30392.23 & 3289.367 & 1.5 & 0.5 & --0.05657 &   0.006123 & 0.12760 & 176 \\
30392.23 & 3289.367 & 2.0 & 1.0 & --0.03402 &   0.003682 & 0.08925 & 171 \\
30392.23 & 3289.367 & 1.0 & 1.0 & --0.09253 &   0.010015 & 0.01785 & 171 \\
30392.23 & 3289.367 & 1.0 & 0.0 &   0.32919 & --0.035629 & 0.03570 & 171 \\
30392.23 & 3289.367 & 4.0 & 3.0 &   0.12828 & --0.013884 & 0.06049 & 173 \\
30392.23 & 3289.367 & 3.0 & 3.0 &   0.12201 & --0.013206 & 0.02614 & 173 \\
30392.23 & 3289.367 & 3.0 & 2.0 & --0.22795 &   0.024673 & 0.02091 & 173 \\
30392.23 & 3289.367 & 2.0 & 3.0 &   0.16844 & --0.018231 & 0.00747 & 173 \\
30392.23 & 3289.367 & 2.0 & 2.0 & --0.18152 &   0.019647 & 0.02614 & 173 \\
30392.23 & 3289.367 & 1.0 & 2.0 & --0.12622 &   0.013661 & 0.02016 & 173 \\
27061.82 & 3694.192 & 0.5 & 0.5 &   0.10060 & --0.013732 & 0.00130 & 168 \\
27061.82 & 3694.192 & 0.5 & 0.5 &   0.07469 & --0.010196 & 0.03040 & 170 \\
27061.82 & 3694.192 & 0.5 & 0.5 &   0.02054 & --0.002804 & 0.21830 & 172 \\
27061.82 & 3694.192 & 0.5 & 0.5 & --0.02200 &   0.003003 & 0.31830 & 174 \\
27061.82 & 3694.192 & 0.5 & 0.5 & --0.06261 &   0.008547 & 0.12760 & 176 \\
27061.82 & 3694.192 & 1.0 & 1.0 & --0.03284 &   0.004483 & 0.07140 & 171 \\
27061.82 & 3694.192 & 1.0 & 0.0 &   0.38888 & --0.053087 & 0.03570 & 171 \\
27061.82 & 3694.192 & 0.0 & 1.0 & --0.10305 &   0.014067 & 0.03570 & 171 \\
27061.82 & 3694.192 & 3.0 & 3.0 &   0.12312 & --0.016807 & 0.04182 & 173 \\
27061.82 & 3694.192 & 3.0 & 2.0 & --0.22685 &   0.030968 & 0.05227 & 173 \\
27061.82 & 3694.192 & 2.0 & 3.0 &   0.18128 & --0.024746 & 0.05227 & 173 \\
27061.82 & 3694.192 & 2.0 & 2.0 & --0.16869 &   0.023029 & 0.01494 & 173 \\
\enddata

\tablenotetext{a}{Center-of-gravity value}

\tablenotetext{b}{Relative to the center-of-gravity value}

\tablenotetext{c}{Normalized to 1 for the whole transition}

\tablecomments{This table is presented as a formatted ASCII file in the electronic edition of the Astrophysical Journal Supplement.}

\end{deluxetable}

\end{center}

\end{document}